\documentclass[prd,preprint,tightenlines,floatfix,showpacs,preprintnumbers,nofootinbib,eqsecnum]{revtex4}
 \usepackage[dvips,final]{graphicx}
  \usepackage{amssymb}
   \usepackage{amsmath}
    \usepackage{epsfig}
     \usepackage{bm}% bold math
      \usepackage{pifont}
\def\agoth{\relax\ifmmode{\mathfrak A}\else{${\mathfrak A}${ }}\fi}
\def\muF{\relax\ifmmode\mu_\text{F}^2\else{$\mu_\text{F}^2${ }}\fi}
\def\muR{\relax\ifmmode\mu_\text{R}^2\else{$\mu_\text{R}^2${ }}\fi}
\def\muO{\relax\ifmmode{\mu_{0}^{2}}\else{$\mu_{0}^{2}${ }}\fi}
\def\Mev{\relax\ifmmode{\text{MeV}}\else{MeV{ }}\fi}
\def\MS{$\overline{\text{MS}\vphantom{^1}}${ }}

\def\Li{\relax\ifmmode{\textbf{Li}_{2}}\else{Li$_2${ }}\fi}

\newcommand{\gev}[1]{\relax\ifmmode{\text{GeV}^{#1}}\else{GeV$^{#1}${ }}\fi}

\def\asb{\relax\ifmmode \bar{\alpha}_s\else{$ \bar{\alpha}_s${ }}\fi}
\def\as{\relax\ifmmode \alpha_s\else{$ \alpha_s${ }}\fi}
\def\acal{\relax\ifmmode{\cal A}\else{${\cal A}${ }}\fi}

\newcommand{\Ds}{\displaystyle}                           %%%%%%%%%

\newcommand{\unl}{\underline}                             %%%%%%%%%
\def\as{\relax\ifmmode \alpha_s\else{$ \alpha_s${ }}\fi}  %%%%%%%%%
\def\abar{\relax\ifmmode{\bar{a}}\else{$\bar{a}${ }}\fi}  %%%%%%%%%
\begin{document}
\thispagestyle{empty}
\date{\today}
\preprint{\hbox{RUB-TPII-03/06}}
%\vspace*{-10mm}

\title{Fractional Analytic Perturbation Theory in Minkowski space
 and application to Higgs boson decay into a $b\bar{b}$ pair \\}
      \author{A.~P.~Bakulev}
\email{bakulev@theor.jinr.ru}
\affiliation{Bogoliubov Laboratory of Theoretical Physics, JINR,
             141980 Dubna, Russia\\}

\author{S.~V.~Mikhailov}%
\email{mikhs@theor.jinr.ru}
\affiliation{Bogoliubov Laboratory of Theoretical Physics, JINR,
             141980 Dubna, Russia\\}

\author{N.~G.~Stefanis}
\email{stefanis@tp2.ruhr-uni-bochum.de}
\affiliation{Institut f\"{u}r Theoretische Physik II,
             Ruhr-Universit\"{a}t Bochum,
             D-44780 Bochum, Germany}
\vspace {10mm}
%\cleardoublepage
\begin{abstract}
We work out and discuss the Minkowski version of Fractional
Analytic Perturbation Theory (MFAPT) for QCD observables, recently
developed and presented by us for the Euclidean region. 
The original analytic approach to QCD, initiated by Shirkov and
Solovtsov, is summarized and relations to other proposals to
achieve an analytic strong coupling are pointed out. 
The developed framework is applied to the Higgs boson decay into 
a $b\bar{b}$ pair, using recent results for the massless correlator 
of two quark scalar currents in the \MS scheme. 
We present calculations for the decay width within MFAPT  
including those non-power-series contributions 
that correspond to the ${\cal O}\left(\alpha_{s}^{3}\right)$-terms,
taking also into account evolution effects of the running coupling 
and the $b$-quark-mass renormalization.
Comparisons with previous results within standard QCD perturbation 
theory are performed and the differences are pointed out.
The interplay between effects originating from the analyticity
requirement and the analytic continuation from the spacelike to
the timelike region and those due to the evolution of the
heavy-quark mass is addressed, highlighting 
the differences from the conventional QCD perturbation theory.
\end{abstract}
%Keywords: General properties of perturbation theory (in QCD)
%          Perturbative calculations in QCD
%          Summation of perturbation theory
%          Renormalization group evolution of parameters
\maketitle

%\tableofcontents

%%%%%%%%%%%%%%%%%%%%%%%%%%%%%%%%%%%%%%%%%%%%%%%%%%%%%%%%%%%%%%%%%%%%%%%
%%%%%%%%%%%%%%%%%%%%%%%%%%%%%%%%%%%%%%%%%%%%%%%%%%%%%%%%%%%%%%%%%%%%%%%
\section{Introduction}
\label{sec:intro}
%%%%%%%%%%%%%%%%%%%%%%%%%%%%%%%%%%%%%%%%%%%%%%%%%%%%%%%%%%%%%%%%%%%%%%%

QCD perturbation theory in the spacelike (Euclidean) domain is based
on a power-series expansion in terms of the running (effective)
coupling $\alpha_{s}(Q^2)$, ($Q^{2}=-q^2>0$), which in one-loop order
reads
\begin{equation}
  \alpha_{s}(Q^2)
=
  \frac{4\pi}{b_0}~a(L)=\frac{4\pi}{b_0}~\frac{1}{L}~
\label{eq:1}
\end{equation}
%Eq (1.1)
with $\Ds ~b_0=11-\frac{2}{3}N_f$, ~$L=\ln(Q^2/\Lambda^2)$, where
$\Lambda^2\equiv \Lambda_\text{QCD}^2$, and with the ``normalized''
coupling $a(L)$ satisfying the renormalization-group equation
\begin{equation}
  \frac{da(L)}{dL}
=
  - a^{2}\left(1 + c_{1} a^{} + c_{2}a^{2}\ldots \right) \, .
\label{eq:2}
\end{equation}
%Eq (1.2)
Here $c_1=b_1/b_0^2$ and $c_2=b_2/b_0^3$ are auxiliary expansion 
parameters (see Appendix \ref{RG-solution} and also Section III in 
Ref.\ \cite{BMS05}).
The one-loop solution of this equation suffers from an artificial
singularity at $L=0$, called the Landau pole.
This prevents the application of perturbative QCD in the low-momentum
spacelike regime with the effect that hadronic quantities, calculated
at the partonic level in terms of a power-series expansion in the
running coupling, are not well defined. 
Besides, from the theoretical point of view, such ghost singularities 
contradict causality rendering a spectral K\"{a}ll\'{e}n--Lehmann 
representation meaningless.
On the other hand, in the timelike, Minkowski, region ($q^{2}>0$) the
definition of the running coupling turns out to be difficult.
The origin of the problem is that the QCD perturbative expansion
cannot be defined in a direct way in this domain.

Many efforts have been made since the early days of QCD to define an   
appropriate coupling parameter in Minkowski space in order to describe
crucial timelike processes like the $e^{+}e^{-}$ annihilation into
hadrons, quarkonium and $\tau$-lepton decays into hadrons, etc.
Most of these attempts---see, for instance,
\cite{PR81,PRR83,Mar88}---were based on the analytic continuation
of the strong coupling from the deep Euclidean region, where
perturbative QCD calculations are safely performed, to the Minkowski
space, where physical measurements are carried out.
Over the years, it became clear that in the infrared (IR), the strong
coupling may reach a stable fixed point and cease to increase.
This behavior would imply that color forces may saturate at this
low-momentum scale meaning that gluons decouple from quarks because
they ``see'' them as a whole, i.e., in a quasi colorless configuration.
Cornwal \cite{Cor82} studied, within a vortex-condensate formalism, the
formation of a mass gap, or effective gluon mass, that prevents the
strong coupling becoming infinite at the Landau pole.
Similar attempts were undertaken by other authors in subsequent years
\cite{PP79,GHN93,MS92} using different techniques, but basing their
arguments on the gluon acquiring an effective mass that works like an
IR regulator in the low-momentum region.
It was shown in \cite{Ste99} that this version of the IR-protected
strong coupling can be related to the Sudakov factor for the
non-emission of soft gluons, in the sense that gluons with wavelengths
above some characteristic (nonperturbative) length scale, cannot
resolve individual quarks because these segregate into a colorless
mock-hadron state.

In separate parallel developments, Radyushkin \cite{Rad82}, and
Krasnikov and Pivovarov \cite{KP82} have obtained analytic expressions
for the one-loop running coupling (and its powers) directly in
Minkowski space using an integral transformation from the spacelike
to the timelike regime reverse to that for the Adler $D$-function (for
more details, we refer the interested reader to 
\cite{Shi00,BRS00,Shi01}).
This sort of analytic coupling in the timelike region was rediscovered
in the context of the resummation of fermion bubbles by Beneke and
Braun \cite{BB95} and also by Ball, Beneke, and Braun in \cite{BBB95},
the latter work in connection with techniques and applications to the
$\tau$ hadronic width.

A systematic approach, termed Analytic Perturbation Theory (APT), has
emerged in the last decade from studies initiated by Shirkov and
Solovtsov \cite{SS96,SS97}.
The main quantity of this framework is the spectral density with the
aid of which an analytic running coupling is defined in the Euclidean
region using a spectral representation.
The same spectral density can be used to define the running coupling
in the timelike region having recourse to the dispersion relation for
the Adler function \cite{MS97,MiSol97}.
These integral transformations, called $\hat{R}$ and $\hat{D}$
operations (see next section), coincide with those invented in
\cite{Rad82,KP82} and provide the possibility to define simultaneously
an analytic running coupling in both the Euclidean and the Minkowski
space.
Meanwhile this analytic approach has been extended beyond the one-loop
level \cite{MiSol97,SS98} and important techniques for numerical
calculations have been developed 
\cite{Mag99,Mag00,KM01,Mag03u,KM03,Mag05}.
The approach has already been applied to the calculation of several
hadronic quantities, important examples being the inclusive decay of a
$\tau$-lepton into hadrons \cite{JS95-349,MSS97,MSSY00}, the
momentum-scale and scheme dependence of the Bjorken \cite{MSS98} and 
the Gross--Llewellyn Smith sum rule \cite{MSS98GLS}, $\Upsilon$ decays 
into hadrons \cite{SZ05}, etc.
Moreover, it has been extended to processes, like the 
$\gamma^*\gamma\to\pi$
transition form factor \cite{SSK99,SSK00} and the pion's
electromagnetic form factor at the next-to-leading order (NLO) of QCD
perturbation theory \cite{SSK99,SSK00,BPSS04}, processes that contain
more than a single perturbative scale, accounting also for Sudakov
suppression.

Overall, this analytic approach (see for some review 
\cite{SS99,Shi00,SS06}) does provide a quite reliable description of 
hadronic quantities in QCD, though there is also criticism \cite{GIZ01} 
and alternative proposals and views of how to avert singularities in 
the running coupling
\cite{Piv91a,LeDiPi92,DMW96,Dok98,CMNT96,Gru97,GGK98,Magn00,%
BRS00,Piv01,GKP01,Nes03,Ale05,CV06,Rac06}---in particular concerning 
the deep IR region $Q^2\leq \Lambda^2$, where eventually the presence 
of a non-vanishing hadronic mass may become important \cite{NP04}.
It also suffers severe limitations in concept and application to
processes beyond the leading order (LO) of QCD perturbation theory
because it assumes that the only quantities that have to be analytic in
the complex $Q^2$ plane are the running coupling and its integer 
powers.
But it was shown in \cite{DR81,BT87,Mul98,MNP99a,MNP01a,MMP02}
that typical three-point functions, like the electromagnetic or
pion-photon transition form factor, at the NLO level of perturbative
QCD, and beyond, contain typical logarithms depending on an additional
scale that serves as a factorization or evolution scale.
These logarithms, though not affecting the Landau singularity, do
contribute to the spectral density.
This has led Karanikas and Stefanis (KS) \cite{KS01,Ste02} to extend
the concept of analyticity (the dispersion relations) from the level 
of the coupling and its powers to the level of QCD hadronic amplitudes 
as a whole.
This generalized encompassing version of the analyticity requirement
demands that all terms that may contribute to the spectral density,
i.e., affect the discontinuity across the cut along the negative real
axis $-\infty<Q^2<0$, must be included into the ``analytization''
procedure.
The implementation of the KS analyticity postulate entails
the extension of the original APT to non-integer (fractional)
powers of the running coupling, which, in particular, encompasses the
logarithms of the factorization (evolution) scale just mentioned.

In this context it is worth emphasizing that fractional powers of
the strong coupling were considered implicitly in
\cite{BKM01}.\footnote{It is interesting to recall here early
attempts to study a spectral density amounting to fractional
indices of the coupling in QED \cite{Shi60}. Such a spectral
density was reinvented later within QCD by Oehme \cite{Oeh90}.}
The systematic development of the Fractional Analytic Perturbation
Theory (FAPT) for QCD in the Euclidean space was recently carried out
in \cite{BMS05} (see also \cite{SBKM06} for a brief introduction) and 
was applied in \cite{BKS05} to the factorized part of the pion's 
electromagnetic form factor that epitomizes three-point functions in 
perturbative QCD.
The pivotal advantage of this scheme is the diminished sensitivity of
the perturbative result on the factorization scale that parallels the
strong renormalization-scheme independence, already established within
APT at the NLO level with respect to the same observable, in the
exhaustive in-depth analysis of \cite{BPSS04} (an abridged version of
which is given in \cite{SBMPS04}---see also
\cite{Ste04mon,AB04pg,AB04ca}).
The aim of the present investigation is to extend this analytic
framework to the Minkowski space twining the two regions, spacelike
and timelike, for any real index and any argument of the couplings and
creating a new calculational paradigm for applications to hadronic
observables in QCD perturbation theory.
The formal discussion of the main characteristics of FAPT in Minkowski
space (for which we use the abbreviation MFAPT) is carried out for
two- and three-loop running-coupling parameters, investigating also the
convergence properties of this type of expansion.
To assess the consequences of MFAPT for observables and elaborate on 
its advantages in detail, it is best to study a quantity at a high-loop 
order of perturbation theory.
To this end, we consider the correlator of two scalar bottom-quark 
currents, whose imaginary part $R_{\text{S}}(s)$ is directly 
proportional to the decay width of a scalar Higgs boson to a 
bottom-antibottom pair.
Considerable progress has been achieved with respect to this quantity
during the last few years mainly thanks to the efforts of Chetyrkin and
collaborators \cite{Che96,ChKS97,BCK05}.
In the present investigation we will provide estimates for
$R_{\text{S}}(s)$ within multiloop MFAPT,
using exclusively the \MS scheme, and compare them with previous 
various results within conventional perturbative QCD up to the order 
${\cal O}\left(\alpha_{s}^{4}\right)$. 
The evolution effects due to the running of the strong coupling
and the heavy-quark mass will be calculated up to three loops, 
borrowing the corresponding four-loop 
expansion coefficient for the Adler function from \cite{BCK05}.
We emphasize in this context that our investigation is mainly meant to 
expose the conceptual advantages of the method, rather than to be used 
as a phenomenological tool for this quantity.

The paper is organized as follows.
In Sec.\ \ref{sec:basics} we provide a mini review of the main 
features of APT, starting with the Euclidean region and completing the 
section with a discussion of the Minkowski domain.
Though this material is mostly based on previously published works, 
its presentation here for any \emph{real} coupling power is new
and has not appeared before in the literature.    
Section \ref{sec:MFAPT} is devoted to the generalization and extension
of FAPT \cite{BMS05} to timelike momenta, giving rise to MFAPT.
In this section we present our main theoretical results and provide the
reader with explicit (but approximate) two-loop expressions for the
analytic powers of the timelike coupling.
Similar formal expressions for any (higher) loop can be derived along 
these lines, with the three-loop case being outlined in 
Appendices \ref{RG-solution}, \ref{app:spectr}, and \ref{app:AnB}).
We close the discussion by giving analysis of the convergence properties
of the perturbative expansion within our analytic scheme.
Section \ref{sec:higgs} contains an application of our framework on
the correlator of two scalar currents of $b$-quarks and the 
Higgs boson decay into a bottom--antibottom pair, presenting estimates 
for the width (actually for the quantity $R_\text{S}$) of this process 
for different orders of the perturbative expansion, as specified above,
and comparing our results with those obtained with the standard
perturbative QCD expansion.
Emphasis is put on the inherent advantage of our method to include into 
the analytic Minkowski couplings the crucial contributions stemming 
from the resummed $\pi^2$ terms owing to the analytic continuation.
Our conclusions are drawn in Sec.\ \ref{sec:concl}, where we also
compile the benchmarks of FAPT in both the Euclidean and the Minkowski
region.
Important technical details are collected in four appendices.

%%%%%%%%%%%%%%%%%%%%%%%%%%%%%%%%%%%%%%%%%%%%%%%%%%%%%%%%%%%%%%%%%%%%%%%
%%%%%%%%%%%%%%%%%%%%%%%%%%%%%%%%%%%%%%%%%%%%%%%%%%%%%%%%%%%%%%%%%%%%%%%
\section{Conceptual essentials of Analytic Perturbation Theory}
\label{sec:basics}
%%%%%%%%%%%%%%%%%%%%%%%%%%%%%%%%%%%%%%%%%%%%%%%%%%%%%%%%%%%%%%%%%%%%%%%

Here we introduce the main theoretical elements of APT, basing our
considerations on \cite{Shi98,SS99}, the aim being to provide a 
comprehensive mini-review of the subject and equip the reader
with all knowledge necessary for the extension to fractional 
powers in Minkowski space in Sec.\ \ref{sec:MFAPT}.
As mentioned in the Introduction, the initial motivation to invent
new couplings was the desire to interrelate the Adler $D$-function,
\begin{eqnarray}
  D(Q^2,\mu^2)
=
  \sum_{n} d_n(Q^2/\mu^2)~a^n(\mu^2)
\stackrel{\mu^2=Q^2}{\longrightarrow} D(Q^2)
=
  \sum_{n} ~d_n~a^n(Q^2)\,,
\label{eq:D}
\end{eqnarray}
%Eq (2.1)
calculable in the Euclidean domain, 
and the quantity
$\Ds R_{e^+e^-}=
 \frac{\sigma(e^+e^-
 \to \text{hadrons})}{\sigma(e^+e^- \to \mu^+\mu^-)}$,
\begin{eqnarray}
  R(s,\mu^2)
=
  \sum_{m} r_m(s/\mu^2)~a^m(\mu^2)
\stackrel{\mu^2=s}{\longrightarrow}
  R(s)= \sum_{m} ~r_m~a^m(s)\ ,
\label{eq:R}
\end{eqnarray}
%Eq (2.2)
which is measured in the Minkowski region.
Both quantities are considered in standard QCD perturbation theory,
demanding that the couplings satisfy the renormalization-group (RG)    
equation.
In minimal-subtraction renormalization schemes the coefficients
$d_n=d_n(1),~r_m=r_m(1)$, entering, respectively, the r.h.s. of
Eqs.\ (\ref{eq:D}) and (\ref{eq:R}), are numerical constants.
The functions $D$ and $R$ can be related to each other via a dispersion
relation without any reference to perturbation theory.
However, employing a perturbative expansion on the l.h.s. of
Eqs.\ (\ref{eq:D}) and (\ref{eq:R}), one obtains, in fact, a relation
between the powers of $\ln(s/\mu^2)$ and $\ln(Q^2/\mu^2)$ in the
coefficients $r_m(s/\mu^2)$ and $d_n(Q^2/\mu^2)$, while the powers of
$\alpha_s(\mu^2)$ reveal themselves as numerical parameters.
Upon setting $\mu^2=Q^2$ on the r.h.s. of Eq.\ (\ref{eq:D}) (or
$\mu^2=s$ in Eq.\ (\ref{eq:R})), the coefficients $d_n$ (analogously
$r_n$) become constants, whereas the coupling powers $\alpha^n_s(Q^2)$
(equivalently, $\alpha^m_s(s)$) are part of the integral
transformations (see below).
But, if these coupling parameters are the standard running ones, then
this connection fails at any loop order because of the Landau
singularity in the Euclidean space.
Questions arise whether analytic versions of both types of couplings
may exist for which the above expressions could be connected.
We shall see in the next step how this goal can, indeed, 
be achieved using non-power-series (functional) expansions 
within APT \cite{SS97,MS97,MiSol97,Shi98,SS99,SS06,Shi00b,Shi00,Shi01}.

The analytic images of the powers of the normalized running coupling, 
cf.\ Eq.\ (\ref{eq:1}), in the Euclidean space can be defined by the
formal linear operation $\textbf{A}_\text{E}$:
\begin{eqnarray}
 \textbf{A}_\text{E}\left[a^{n}_{(l)}\right]
  &=& {\cal A}^{(l)}_{n}
    \text{~~~with~~~}
   {\cal A}^{(l)}_{n}(Q^2)
  \equiv
   \int_0^{\infty}\!
    \frac{\rho^{(l)}_n(\sigma)}
         {\sigma+Q^2}\,
       d\sigma\,,
\label{eq:A.E}
\end{eqnarray}
%Eq (2.3)
where the spectral density is defined as
\begin{eqnarray}
\label{eq:rho}
\rho^{(l)}_n(\sigma)
   \equiv \frac{1}{\pi}\,
           \textbf{Im}\,\big[a^{n}_{(l)}(-\sigma)\big]\,.
\end{eqnarray}
%Eq (2.4)
The power (index) $n$ denotes here only integer values, whereas the 
loop order is indicated by $l$ in parenthesis.
We will show later that these relations are valid also for fractional
powers (indices) $\nu$.

\begin{figure}[t]
 $$\includegraphics[width=0.3\textwidth]{%fig-Cont.eps
 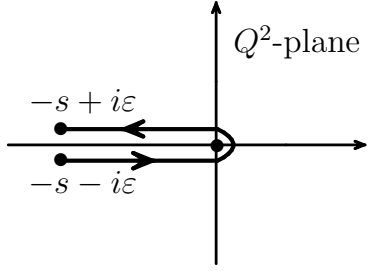}$$
 \vspace{-0.3cm}
 \caption{Integration contour for the $\hat{R}$-operation in
 Eq.\ (\protect\ref{eq:R-operation}).
 \label{fig:contour}}
\end{figure}

Analogously, the analytic images of the normalized running coupling
in Minkowski space are defined by means of another linear operation,
$\textbf{A}_\text{M}$, viz.,
\begin{eqnarray}
 \textbf{A}_\text{M}\left[a^{n}_{(l)}\right]
  = {\mathfrak A}^{(l)}_{n}
  \text{~~~with~~~}
  {\mathfrak A}^{(l)}_{n}(s)
  \equiv
    \int_s^{\infty}\!
     \frac{\rho^{(l)}_n(\sigma)}
          {\sigma}\,
      d\sigma\,.
\label{eq:A.M.rho}
\end{eqnarray}
%Eq (2.5)
These ``analytization'' operations can be represented by the following
two integral transformations:
\begin{itemize}
\item $\hat{D}$ from the timelike region to the spacelike region
\begin{eqnarray}
  \hat{D}\big[{\mathfrak A}^{(l)}_{n}\big]
  &=&  {\cal A}^{(l)}_{n}
    \text{~~~with~~~}
   {\cal A}^{(l)}_{n}(Q^2)
  \equiv
   Q^2 \int_0^{\infty}\!
     \frac{{\mathfrak A}^{(l)}_{n}(\sigma)}
          {\big(\sigma+Q^2\big)^2}\,
      d\sigma
\label{eq:D-operation}
\end{eqnarray}
%Eq (2.6)
and \item $\hat{R}$ for the inverse transformation (adopting the
terminology of Shirkov, see, for instance, \cite{Shi00,Shi01,Shi05})
\begin{eqnarray}
  \hat{R}\big[{\cal A}^{(l)}_{n}\big]
  &=&
  {\mathfrak A}^{(l)}_{n}
  \text{~~~with~~~}
  {\mathfrak A}^{(l)}_{n}(s)
  \equiv \frac{1}{2\pi i}
   \int_{-s-i\varepsilon}^{-s+i\varepsilon}\!
    \frac{{\cal A}^{(l)}_{n}(\sigma)}
          {\sigma}\,
      d\sigma\,,
\label{eq:R-operation}
\end{eqnarray}
%Eq (2.7)
\end{itemize}
where the last integral is evaluated along the contour shown in
Fig.\ \ref{fig:contour}.
%%%%%%%%%%%%%%%%%%%%%%%%%%%%%%FIGURE 1%%%%%%%%%%%%%%%%%%%%%%%%%%%%%%%%%
%%%%%%%%%%%%%%%%%%%%%%%%%%%%%%%%%%%%%%%%%%%%%%%%%%%%%%%%%%%%%%%%%%%%%%%
Note that these operations are connected to each other by the relation
\begin{eqnarray}
 \label{eq:reciproc}
 \hat{D}\hat{R} = \hat{R}\hat{D} = 1\,,
\end{eqnarray}
%Eq (2.8)
valid for the whole set
$\big\{{\cal A}_n,{\mathfrak A}_n\big\}$ and at any loop order.

\begin{figure}[t]
 \centerline{\includegraphics[width=0.33\textwidth]{%fig-APT.eps
  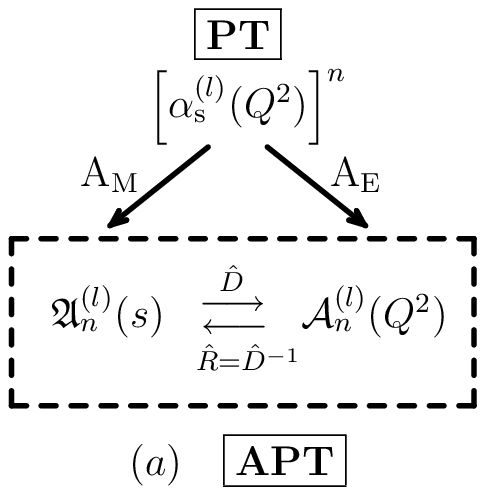}}
  \vspace*{2mm}
 \centerline{\includegraphics[width=0.33\textwidth]{%fig-FAPT.eps
 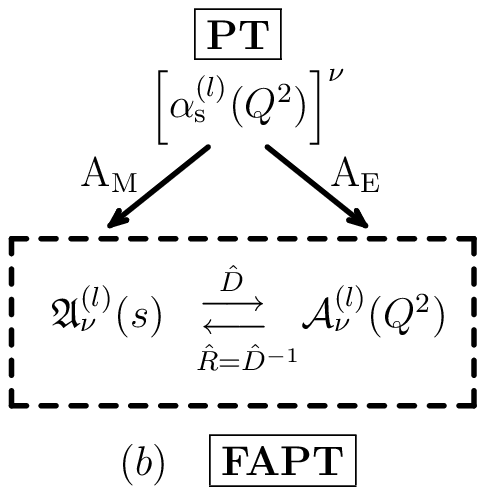}
 ~~~\includegraphics[width=0.33\textwidth]{%fig-FAPT2.eps
 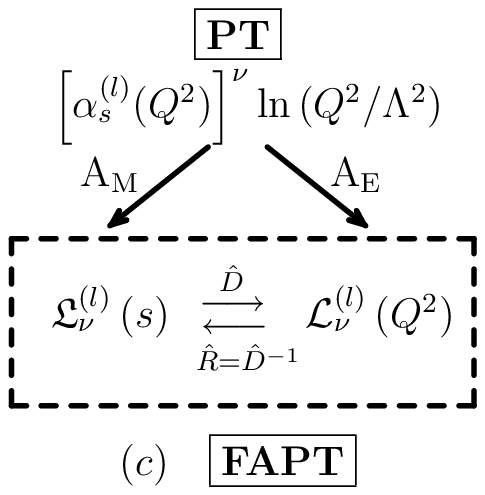}}
   \vspace{0.0cm}
   \caption{Implementation of analyticity in APT (a) and in FAPT (b)
   and (c).
   The index $n$ in APT is restricted to integer values only, while in
   FAPT $\nu$ can assume any real value, enabling the ``analytization''
   of expressions like those shown in (c) and presented in Appendix
   \protect\ref{app:AnB}.
   \label{fig:APT-scheme}}
\end{figure}

The operations $\textbf{A}_\text{E}$ and $\textbf{A}_\text{M}$, which 
define, respectively, the analytic running couplings in the Euclidean 
(spacelike) and in the Minkowski (timelike) region are displayed 
graphically in Fig.\ \ref{fig:APT-scheme}.
The logic of ``analytization'' enables a similar outcome with respect 
to the expansion of QCD amplitudes (depending on a single momentum 
scale $Q^2$) and their continuation from the Euclidean to the 
Minkowski space.
As an example, consider the Adler function $D$ on the r.h.s. of 
Eq.\ (\ref{eq:D}), which is expanded in terms of $\alpha^n_s(Q^2)$.
The operation $\textbf{A}_\text{E}$, 
applied to $D(Q^2)$ along the right arrow in Fig.\ \ref{fig:APT-scheme}(a), 
maps it on a non-power-series expansion \cite{Shi98,Shi00} 
in the Euclidean region, termed ${\cal D}_\text{A}$, i.e.,
\begin{eqnarray}
  D(Q^2)
  = \sum_{n}\,d_n\,a_{(l)}^n(Q^2)\, \Rightarrow
     \textbf{A}_\text{E}\left[D \right]
  \equiv   {\cal D}_\text{A}
    \text{~~~~~with~~~~~}
  {\cal D}_\text{A}(Q^2)
  = \sum_{n}\,d_n\,{\cal A}^{(l)}_{n}(Q^2)\, .
\label{eq:cal D}
\end{eqnarray}
%Eq (2.9)
%%%%%%%%%%%%%%%%%%%%%%%%%%%%%%FIGURE 2%%%%%%%%%%%%%%%%%%%%%%%%%%%%%%%%%
%%%%%%%%%%%%%%%%%%%%%%%%%%%%%%%%%%%%%%%%%%%%%%%%%%%%%%%%%%%%%%%%%%%%%%%
Subsequently, one can apply the $\hat R$ operation, given by Eq.\
(\ref{eq:R-operation}) (bottom line in Fig.\ \ref{fig:APT-scheme}(a)),
to obtain the quantity $R$ in the Minkowski region:
\begin{eqnarray}
  \hat{R}\left[{\cal D}_\text{A} \right]
  \equiv {\cal R}
     \text{~~~~~with~~~~~}
  {\cal R}(s)
  = \sum_{n}\,d_n\,{\mathfrak A}^{}_{n}(s)\, .
\label{eq:cal R}
\end{eqnarray}
%Eq (2.10)
On the other hand, the same expression for ${\cal R}(s)$ (in Minkowski
space) can be obtained following the left arrow in Fig.\
\ref{fig:APT-scheme}(a) by making use of the $\textbf{A}_\text{M}$
operation.
This leads to the same analytic image of $R(s)$:
\begin{eqnarray}
 D(Q^2)
 = \sum_{n}\,d_n\,a_{(l)}^n(Q^2)\,
 \Rightarrow
  \textbf{A}_\text{M}\left[D\right]
 = {\cal R}
     \text{~~~~~with~~~~~}
  {\cal R}(s)
 = \sum_{n}\,d_n\,{\mathfrak A}^{}_{n}(s)\,.
\label{eq: DD}
\end{eqnarray}
%Eq (2.11)
The generalized concept of imposing analyticity to the QCD amplitude as
a whole \cite{KS01,Ste02}, allows us to perform the ``analytization''
of any fractional (real) power of the coupling and---even more
important---invoke the ``analytization'' concept on more complicated 
expressions that contain products of powers of the strong coupling 
times logarithms of a second perturbative scale, like the factorization 
or evolution scale---see for an illustration part (c) of 
Fig.\ \ref{fig:APT-scheme} and for details Appendix \ref{app:AnB}.
Then, all of the previous results, exposed via Eqs.\
(\ref{eq:A.E})--(\ref{eq:reciproc}), can be generalized to hold for any
fractional index (power) $\nu$, giving rise to the vector spaces
$\big\{{\cal A}_\nu\big\}, \big\{{\mathfrak A}_\nu\big\}$ that possess
the property of index differentiation.

Let us now turn our attention to the spectral density.
At the one-loop level, we can derive by a straightforward calculation,
based on Eqs.\ (\ref{eq:1}) and (\ref{eq:rho}), a closed-form
expression for the spectral density; namely,\footnote{An analogous 
expression for QED can be found in \cite{Shi60}.}
\begin{eqnarray}
 \rho_{\nu}^{(1)}(\sigma)
  = \frac{1}{\pi}\,
   \frac{\sin(\nu~\varphi_\sigma) }{\left[\pi^2+L^2_\sigma\right]^{\nu/2}}
 \,,\quad
 \varphi_\sigma
  = \arccos\left(\frac{L_{\sigma}}{\sqrt{L^2_\sigma+\pi^2}}\right)
 \,,\quad
 L_{\sigma}
  = \ln\left(\sigma/\Lambda^2\right)\,.
\label{eq:rho-A-dis}
\end{eqnarray}
%Eq (2.12)
Then, the one-loop couplings
${\cal A}^{(1)}_{1}$ and ${\mathfrak A}^{(1)}_{1}$
can be derived by substituting $\rho_{1}^{(1)}$ into
Eqs.\ (\ref{eq:A.E}) and (\ref{eq:A.M.rho}) to get \cite{SS97}
\begin{eqnarray}
  {\cal A}^{(1)}_{1}(Q^2)
  &=&\frac1{L}- \frac1{e^L -1}\,,
 \label{eq:A_1}
\end{eqnarray}
%Eq (2.13)
and \cite{MiSol97}
\begin{eqnarray}
{\mathfrak A}^{(1)}_{1}(s)
  &=& \frac{1}{\pi}\,
       \arccos\left(\frac{L_{s}}{\sqrt{L_{s}^2+\pi^2}}\right)
 \label{eq:U_1}
\end{eqnarray}
%Eq(2.14)
with
\begin{eqnarray}
 \label{eq:L.s.L.sigma}
  L \,=\, \ln\left(Q^2/\Lambda^2\right)\,,\qquad
  L_s \,=\, \ln\left(s/\Lambda^2\right)\,.
\end{eqnarray}
%Eq (2.15)

From these equations we infer that ``analytization''
($\textbf{A}_\text{E}$) in the Euclidean case amounts to the
subtraction (at the one-loop level) of the Landau pole, whereas in
the Minkowski space the analogous operation ($\textbf{A}_\text{M}$)
means summation of $\pi^2$-terms in all orders of the expansion.
To generate two-loop expressions for the analytic couplings,
one can make use of the Lambert function, as shown by Magradze in
\cite{Mag99}.
Still higher loops can be obtained via an approximate form of the
spectral density and numerical integration \cite{Shi98}.
It follows from this summarized exposition that the difficulties
encountered with ghost singularities in the Euclidean strong
coupling can be eliminated on account of causality (``spectrality''
\cite{Shi98}) and RG invariance.
Hence, from the point of view of the analytic approach, the Landau
pole remover (and analogously the compensation of singularities in
higher loops) is not introduced by hand but ensues naturally as a
corollary within the formalism without appealing to any
nonperturbative physics as the origin of power corrections.
Nevertheless, one may include into the spectral density power
corrections of the form $\left(M^2/Q^2\right)^n$, with $M$ being, for
example, a constituent quark mass, 
in an attempt to incorporate this way some nonperturbative effects.
Such ``analytization'' approaches, following, however, different
incentives, have been proposed in \cite{CV06,Ale05}.

%%%%%%%%%%%%%%%%%%%%%%%%%%%%%%%%%%%%%%%%%%%%%%%%%%%%%%%%%%%%%%%%%%%%%%%
%%%%%%%%%%%%%%%%%%%%%%%%%%%%%%%%%%%%%%%%%%%%%%%%%%%%%%%%%%%%%%%%%%%%%%%
\section{Minkowski version of Fractional Analytic Perturbation Theory}
\label{sec:MFAPT}
%%%%%%%%%%%%%%%%%%%%%%%%%%%%%%%%%%%%%%%%%%%%%%%%%%%%%%%%%%%%%%%%%%%%%%%

Before studying the detailed procedure of the analytic continuation to
Minkowski space, let us first make some important remarks about the
strategy on how to generalize the approach in order to include
non-integer (fractional) indices $\nu$.
Appealing to our detailed discussion in \cite{BMS05}, we note that it 
is not obvious how the direct way for the analytic continuation of the 
coupling powers in terms of Eq.\ (\ref{eq:A.E}) can provide explicit 
analytic expressions.
Therefore, to achieve this goal we have used instead in \cite{BMS05} 
another method, based on the Laplace representation, which will be 
exposed in the next subsection.
However, the situation in Minkowski space is different because of the
absence of ghost singularities.
In that case it turns out to be possible to employ the
dispersion-relation techniques (cf.\ Eq.\ (\ref{eq:A.M.rho})) in order
to obtain analytic expressions in the timelike regime that are valid
for any real index $\nu$.
Indeed, using Eq.\ (\ref{eq:rho-A-dis}), we find in one-loop order
\begin{eqnarray}
 {\mathfrak A}^{(1)}_{\nu}(s)
  & = &
  \int_{s}^{\infty} \frac{d\sigma}{\sigma}\,
   \rho^{(1)}_{\nu}(\sigma)
  = \frac{1}{\pi}\, \int_{L_s}^{\infty}\!\!
     dL\,
     \frac{\sin\left[\nu\arccos\left(L/\sqrt{(L^2+\pi^2)}\right)
               \right]}
          {\left(\pi^{2} + L^{2}\right)^{\nu/2}}\,.
\label{eq:timelike}
\end{eqnarray}
%Eq (3.1)
This integral can be evaluated explicitly and provides the result
 \begin{eqnarray}
 {\mathfrak A}^{(1)}_{\nu}(s)
  & = & \frac{\sin\left[(\nu -1)\arccos\left(L_s/\sqrt{(L_s^2+\pi^2)}
                                       \right)
                  \right]}
             {\pi\,(\nu -1)\,\left(L_s^2+\pi ^2\right)^{(\nu-1)/2}}
\label{eq:TL_ElFun}
\end{eqnarray}
%Eq (3.2)
that is completely determined by elementary functions
\cite{BMS05,SBKM06}.
Taking the limit $\nu\to 1$ in the above equation, one readily obtains
Eq.\ (\ref{eq:U_1}) for ${\mathfrak A}^{(1)}_{1}$, while taking
$\nu=0$ one finds ${\mathfrak A}^{(1)}_{0}=1$.

Let us consider now the spectral density $\rho_{\nu}^{(l)}(\sigma)$
beyond the leading-order approximation.
At the $l$-loop level, $\rho_{\nu}^{(l)}(\sigma)$ can always be
presented in the same form as for the leading-order one, given by
Eq.\ (\ref{eq:rho-A-dis}), i.e.,
\begin{eqnarray}
  \rho_{\nu}^{(l)}(\sigma)
=
  \frac{1}{\pi}\,
  \textbf{Im}\,\big[a^{\nu}_{(l)}(-\sigma)\big]
= \frac{\sin[\nu\,\varphi_{(l)}(\sigma)]}
       {\pi\,\left(R_{(l)}(\sigma)\right)^{\nu}}
\label{eq:spec-dens-n}
\end{eqnarray}
%Eq (3.3)
keeping, however, in mind that the phase $\varphi_{(l)}$ and the radial
part $R_{(l)}$ acquire now a multi-loop content.
Suffice it to mention here that an explicit two-loop expression for the 
spectral density is derived in Appendix \ref{app:spectr}, notably, 
Eq.\ (\ref{eq:rho2-app}). 
In the same appendix we show that this expression, though approximate,
is very close to the exact, but numerical, one. 
To be more specific, one should, strictly speaking, deal with the 
imaginary part of the appropriate branch of the Lambert function 
$W_{-1}$ (see \cite{Mag99}) owing to the fact that the exact solution 
of the two-loop RG equation (given in Appendix \ref{RG-solution} by 
Eqs.\ (\ref{eq:App-RGExact}) and (\ref{eq:App-Exactsolution})) can be 
expressed in terms of this function.
To complete the exposed procedure, one should substitute the displayed
spectral density into Eq.\ (\ref{eq:A.M.rho}) and perform the
integration.
Recently, Magradze \cite{Mag00,Mag05} has published closed-form
expressions for ${\mathfrak A}_{1}^{(2)}$, ${\mathfrak A}_{2}^{(2)}$ at
the two-loop level by means of the $W_{1}$ Lambert function.
The dark side of this latter procedure is that it does not lend itself
to an analytic evaluation of explicit expressions for
${\mathfrak A}_{\nu}^{(2)}(L_s)$ for \emph{fractional} indices, but 
yields (after integration) only numerical values.
Beyond the two-loop level, explicit results are difficult to obtain.
Numerical values of the quantities
${\cal A}_{n}$ and ${\mathfrak A}_{n}$ for $n=1,2,3$
at the three-loop level were given by Kourashev and Magradze in
\cite{KM01}.
Very recently, Shirkov and Zayakin \cite{SZ05} have constructed 
a simple one-parameter model 
to emulate the first three ($n=1,2,3$) analytic 
couplings at the three-loop level, both in the Euclidean and the 
Minkowski region, that claims an acceptable accuracy for practical 
purposes.

%%%%%%%%%%%%%%%%%%%%%%%%%%%%%%%%%%%%%%%%%%%%%%%%%%%%%%%%%%%%%%%%%%%%%%%
\subsection{Simultaneous derivation of FAPT in the Euclidean and
            Minkowski regions at the one-loop level}
\label{sec:laplace}
%%%%%%%%%%%%%%%%%%%%%%%%%%%%%%%%%%%%%%%%%%%%%%%%%%%%%%%%%%%%%%%%%%%%%%%

In this subsection, we consider  timelike and spacelike couplings in 
mutual comparison, exclusively in the one-loop approximation, omitting 
for this reason the loop label $(l)$.
The generalization to higher loops will be presented in Subsection 
\ref{subsec:convergence}.
Let us start the derivation of the (M)FAPT analytic couplings
$\{{\cal A}_{\nu}(k),{\mathfrak A}_{\nu}(k)\}$
with $k$ being a logarithm of a momentum (or energy) scale by employing 
the Laplace-representation approach of \cite{BMS05}.
It is useful to recall at this point that the initial sets
$\{{\cal A}_{n}\}$, $\{{\mathfrak A}_{n}\}$ have been constructed for
integer values of the index and constitute vector spaces \cite{BMS05}.
Being able to create the elements
${\cal A}_{\nu}$, ${\mathfrak A}_{\nu}$ for any real $\nu$,
one can complete the vector spaces
$\{{\cal A}_{\nu}\}$, $\{{\mathfrak A}_{\nu}\}$, making it possible
to apply other linear operations to these spaces, e.g., 
differentiation with respect to the index $\nu$.

It is important to appreciate that the generalization of all APT
couplings to fractional (real) values can be performed 
within a single mould.
To this end, we apply a differential-type RG-equation, like
(\ref{eq:2}), and, following \cite{Shi98,BMS05}, we first write
\begin{equation}
\left(
\begin{array}{l}
  a^{n}(k)\\
{\cal A}_{n}(k)\\
 {\mathfrak A}_{n}(k)
 \end{array}
 \right)
 =
   \frac{1}{(n-1)!}\left( -\frac{d}{d k}\right)^{n-1}
\left(
\begin{array}{l}
  a^{1}(k)\\
{\cal A}_{1}(k)\\
 {\mathfrak A}_{1}(k)
 \end{array}
 \right)\, ,
\label{eq:generator}
\end{equation}
%Eq (3.4)
where the evaluation of the couplings $a^{n}$ (standard, $n$: power)
and ${\cal A}_{n}$ (analytic, $n$: index) in the spacelike region
proceeds with $k=L\equiv \ln(Q^2/\Lambda^2)$, while their
counterparts in the timelike region, ${\mathfrak A}_{n}$,
are calculated with the aid of $k= L_s\equiv\ln(s/\Lambda^2)$.

\begin{figure}[h]
 \centerline{\includegraphics[width=0.5\textwidth]{%fig-laplace-1m.eps
 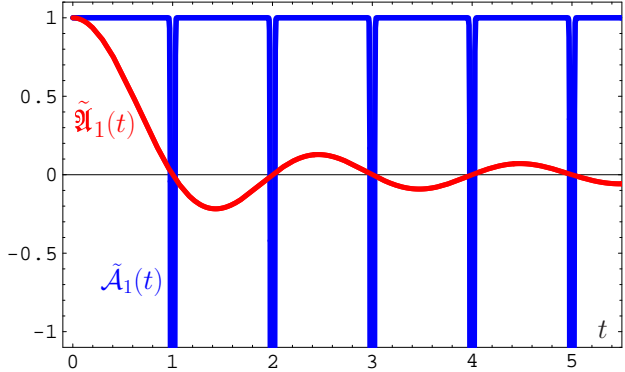}}
   \vspace{0.0cm}
   \caption{Illustration of the Laplace images of the one-loop
   analytic couplings in the Euclidean
   $(\tilde{{\cal A}}_{1}(t))$ and in the Minkowski space
   $(\tilde{\mathfrak A}_{1}(t))$.
   \label{fig:laplace}}
\end{figure}

To facilitate the transition to fractional index values, it is
instrumental to employ the Laplace representation of both types
of couplings---the analytic, ${\cal A}_{n}(l),~{\mathfrak A}_{n}(l)$,
and the conventional ones, $a^n(L)$,---and define for $k > 0$
\begin{equation}
\left(
  \begin{array}{l}
  a^{n}(k)\\
{\cal A}_{n}(k)\\
 {\mathfrak A}_{n}(k)
 \end{array}
 \right)
=
   \int_0^{\infty} e^{-k t}
\left(
 \begin{array}{l}
  \tilde{a}_{n}(t)   \\
      \tilde{{\cal A}}_{n}(t)\\
   \tilde{\mathfrak A}_{n}(t)
 \end{array}
 \right)
    dt \,
 =
   \int_0^{\infty} e^{-k t}
   \frac{t^{n-1}}{\Gamma(n)}
\left(
 \begin{array}{l}
  \tilde{a}_{1}(t)   \\
      \tilde{{\cal A}}_{1}(t)\\
   \tilde{\mathfrak A}_{1}(t)
 \end{array}
 \right) dt \, .
\label{eq:laplace}
\end{equation}
%Eq (3.5)
To derive the last equation, we have used in Eq.\ (\ref{eq:generator})
the one-loop RG equation.
The key element in converting the APT couplings to the set
$\{{\cal A}_{n}(k),~{\mathfrak A}_{n}(k)\}$,
valid for any index $\nu$, is the relation
\begin{equation}
\left(
 \begin{array}{l}
  \tilde{a}_{\nu}(t)   \\
      \tilde{{\cal A}}_{\nu}(t)\\
   \tilde{\mathfrak A}_{\nu}(t)
 \end{array}
 \right)
   \stackrel{\text{def}}=
     \frac{t^{\nu-1}}{\Gamma(\nu)}
\left(
 \begin{array}{l}
  \tilde{a}_{1}(t)   \\
      \tilde{{\cal A}}_{1}(t)\\
   \tilde{\mathfrak A}_{1}(t)
 \end{array}
 \right)
\label{eq:laplace-n}
\end{equation}
%Eq (3.6)
that generalizes Eq.\ (\ref{eq:laplace}).
From this, it is evident that $\tilde{a}_{1}(t)=1$.
The explicit expression for $\tilde{{\cal A}}_{1}(t)$---worked out 
before in \cite{BMS05}---and that for the timelike coupling
$\tilde{\mathfrak A}_{1}(t)$ can be written as follows
\begin{equation}
\left(
 \begin{array}{l}
  \vphantom{^{\big|}} \tilde{a}_{1}(t)   \\
   \vphantom{^{\big|}}   \tilde{{\cal A}}_{1}(t)\\
    \vphantom{^{\big|}} \tilde{\mathfrak A}_{1}(t)
 \end{array}
 \right)
   =
     \frac{t^{\nu-1}}{\Gamma(\nu)}
\left(
 \begin{array}{l}
    \vphantom{^{\big|}} 1   \\
     \vphantom{^{\big|}} 1 - \sum_{m=1}^{\infty} \delta(t-m)\, \\
  \Ds  \vphantom{^{\big|}} \sin(\pi t)/(\pi t)
 \end{array}
 \right) \,.
\label{eq:laplace-1}
\end{equation}
%Eq (3.7)
Note that the first term in the Euclidean analytic coupling (second 
line in the above equations) stems from the usual QCD term $1/L$, 
whereas the $\delta$-function term is related to the Landau-pole 
remover (second term in Eq.\ (\ref{eq:A_1})).
To reveal the particular features of the Laplace-conjugate images of 
these couplings, we show them graphically in Fig.\ \ref{fig:laplace}.
One sees from this figure that the Laplace conjugate of the
conventional (normalized) coupling corresponds to a straight line at 
unity, while the analogous expression for the Euclidean coupling is 
represented by a Dirac comb (blue line) and the Minkowski one is a 
smooth and oscillating function (red line) dying out with $t$.

%%%%%%%%%%%%%%%%%%%%%%%%%%%%%%FIGURE 3%%%%%%%%%%%%%%%%%%%%%%%%%%%%%%%%%
%%%%%%%%%%%%%%%%%%%%%%%%%%%%%%%%%%%%%%%%%%%%%%%%%%%%%%%%%%%%%%%%%%%%%%%
Though we have initially assumed that $k>0$ and $\nu >0$, these Laplace
conjugates generate, in turn, the following images that can be
analytically extended to any real $\nu,~L$, and $L_\sigma$
(we display again the loop label ($l=1$) explicitly for the sake of 
comparison later on)
\begin{eqnarray}
a^{\nu}_{(1)} &=& \frac1{L^\nu}\, , \\
 {\cal A }^{(1)}_{\nu}(L)
          &=& \frac1{L^{\nu}}
   - \frac{F(e^{-L},1-\nu)}{\Gamma(\nu)}\,,
 \label{eq:A-F} \\
{\mathfrak A}^{(1)}_{\nu}(L_s)
  & = & \frac{\sin\left[(\nu -1)\arccos\left(L_s/\sqrt{(L_s^2+\pi^2)}
                                       \right)
                  \right]}
             {\pi\,(\nu -1) \left(L_s^2+\pi ^2\right)^{(\nu-1)/2}}\, ,
  \label{eq:U-F}
\end{eqnarray}
%Eqs (3.8) (3.9) (3.10)
where the Dirac comb gives rise to the transcendental Lerch function,
$F(z,\nu)$ \cite{BE53}, that serves as a Landau-pole remover for any 
$\nu$, and the oscillating curve amounts to elementary functions, 
confirming the result given in Eq.\ (\ref{eq:TL_ElFun}).
The last expression is a new result of the present analysis and has 
been derived by taking recourse to the last entry of 
Eq.\ (\ref{eq:laplace-1}).
It is worth emphasizing that ${\cal A }^{(1)}_{\nu}(L)$ and
${\mathfrak A}^{(1)}_{\nu}(L_s)$ are entire functions of their
corresponding arguments.

%%%%%%%%%%%%%%%%%%%%%%%%%%%%%%%%%%%%%%%%%%%%%%%%%%%%%%%%%%%%%%%%%%%%%%%
\subsection{Properties of timelike vs.\ spacelike couplings for any 
            real index $\nu$}
\label{sec:real-index}
%%%%%%%%%%%%%%%%%%%%%%%%%%%%%%%%%%%%%%%%%%%%%%%%%%%%%%%%%%%%%%%%%%%%%%%

\textbf{1.} From inspection of the relations (\ref{eq:D-operation}),
(\ref{eq:R-operation}), and recalling that asymptotically as
$L\to \infty$ both analytic couplings,
${\cal A}_{\nu}(L)$ and ${\mathfrak A}_{\nu}(L_s)$,
tend to the \emph{same} standard coupling $a^{\nu}(L)$, one may ask
about the mutual behavior of these couplings for finite values of
their arguments.
Despite the asymptotic symmetry of the couplings, for finite arguments
this symmetry is distorted \cite{MS97,MiSol97}, albeit the couplings
are equal at the origin, i.e., for $Q^2=s \to 0$, or equivalently,
$L=L_s \to -\infty$.
This ``distorting-mirror'' effect \cite{MS97,MiSol97} is an interesting
property of the analytic couplings and was originally
established for integer powers of the coupling. 
It is symbolically expressed through the operation
\begin{equation}
 {\cal A}^{(l)}_{\nu}(Q^2)
=
 \hat{D}\big[{\mathfrak A}^{(l)}_{\nu}(s)\big]
\label{eq:distortion}
\end{equation}
%Eq (3.11)
(see Eq.\ (\ref{eq:D-operation})), which we now generalize to be valid
for any real index $\nu$.
Its content may become evident from the following expressions for
$L=L_s=0$:
\begin{eqnarray}
  {\mathfrak A}_{\nu}(0)
=
  \frac{\sin\left[(\nu-1)\pi/2\right]}{(\nu-1)\,\pi^{\nu}}:&&
  {\mathfrak A}_{1}(0)=\frac1{2},~{\mathfrak A}_{2}(0)
=
  \frac1{\pi^2},~{\mathfrak A}_{3}(0)=0,
  ~{\mathfrak A}_{4}(0)=-\frac1{3\pi^4}, ~~~~\\
  &&\Ds {\mathfrak A}_{\frac{1}{2}}(0)=\frac{\sqrt{2}}{\pi^{1/2}},
  ~~~{\mathfrak A}_{\frac{5}{2}}(0)=\frac{\sqrt{2}}{3 \pi^{5/2}},
  ~{\mathfrak A}_{\frac{7}{2}}(0)=-\frac{\sqrt{2}}{5 \pi^{7/2}}\\
  {\cal A}_{\nu}(0)=-\frac{\zeta(1-\nu)}{\Gamma(\nu)}:&&
  {\cal A}_{1}(0)=\frac1{2},~{\cal A}_{2}(0)=
  \frac1{12},~{\cal A}_{3}(0)= 0,~{\cal A}_{4}(0)=-\frac1{720}\,\\
  &&\Ds{\cal A}_{\frac{1}{2}}(0)=\frac{-\zeta(\frac{1}{2})}{\pi^{1/2}},
  ~{\cal A}_{\frac{5}{2}}(0)=\frac{\zeta(\frac{5}{2})}{4 \pi^{5/2}},
  {\cal A}_{\frac{7}{2}}(0)=-\frac{\zeta(\frac{7}{2})}{8 \pi^{7/2}}, 
\end{eqnarray}
%Eqs (3.12) (3.13) (3.14 (3.15)
where $\zeta(\nu)$ is the Riemann $\zeta$ function.
These couplings are interrelated by the equation
\begin{eqnarray}
{\cal A}_{\nu}(0)
=
 \left[\frac{(\nu-1)\,\zeta(\nu)}{2^{\,\nu-1}}
 \right]
 {\mathfrak A}_{\nu}(0)
\end{eqnarray}
%Eq (3.16)
with the coefficient in the bracket providing a quantitative measure 
for the magnitude of the distortion for any $\nu \in \mathbb{R}$.
For a graphic illustration of the ``distorted mirror symmetry'' effect, 
we refer the interested reader to \cite{MS97,Shi01,Shi05}.

\textbf{2.} As we have shown in \cite{BMS05} for the Euclidean
couplings, the parameters ${\cal A }_{-\nu}$ play the role of the
``inverse powers'' of ${\cal A }_{1}$ that may be considered as
the images of $a_s^{-\nu}$.
This property extends also to the Minkowski region, so that the set
$\big\{{\cal A}_{-\nu},{\mathfrak A}_{-\nu}\big\}$
corresponds to analytic images of $a^{-\nu}$ in the Euclidean and the
Minkowski space, respectively, for \emph{arbitrary} $\nu$ values.
This allows us to demonstrate the ``distorted mirror'' effect in
analytic form as follows
\begin{eqnarray}
 \label{U-A-n}
 \begin{array}{llll}
  {\mathfrak A}_{0}(L_{s}) = 1\,,
   & {\mathfrak A}_{-1}(L_{s}) = L_{s}\,,
    & {\mathfrak A}_{-2}(L_{s}) = L_{s}^2-\pi^2/3\,,
     & {\mathfrak A}_{-3}(L_{s}) = L_{s}(L_{s}^2-\pi^2)\,,\ldots \\
  {\cal A}_{0}(L) = 1\,,
   & {\cal A}_{-1}(L) = L\,,
    & {\cal A}_{-2}(L) = L^2\,,
     & {\cal A}_{-3}(L) = L^3\,,\ldots
 \end{array}
\end{eqnarray}
%Eq (3.17)
Note that the expressions for ${\mathfrak A}_{-n}$, given by
Eq.\ (\ref{U-A-n}), can be linked to ${\cal A}_{-n}$, 
in analogy to (\ref{eq:distortion}), by means of the
transformation
\begin{equation}
 {\cal A}^{}_{-n}(Q^2)
=
 \hat{D}\big[{\mathfrak A}^{}_{-n}(s)\big]\, .
\label{eq:E-M-transfor}
\end{equation}
%Eq (3.18)
[See Eq.\ (19) in \cite{ChKS97} for an earlier implicit derivation of
these expressions, employing transformation (\ref{eq:D-operation})
in terms of the powers of $L_s$ (or in terms of
${\mathfrak A}^{}_{-n}(s)$ in our notation)].
A useful all-order formula to analytically continue logarithms
under the $\hat{D}$ transformation (see item \textbf{3} in Appendix
\ref{app:spectr}) has been presented in \cite{BKM01}.

%%%%%%%%%%%%%%%%%%%%%%%%%%%%%%FIGURE 4%%%%%%%%%%%%%%%%%%%%%%%%%%%%%%%%%
\begin{figure}[h]
 \centerline{\includegraphics[width=0.45\textwidth]{%fig_unu23_fapt-b.eps
 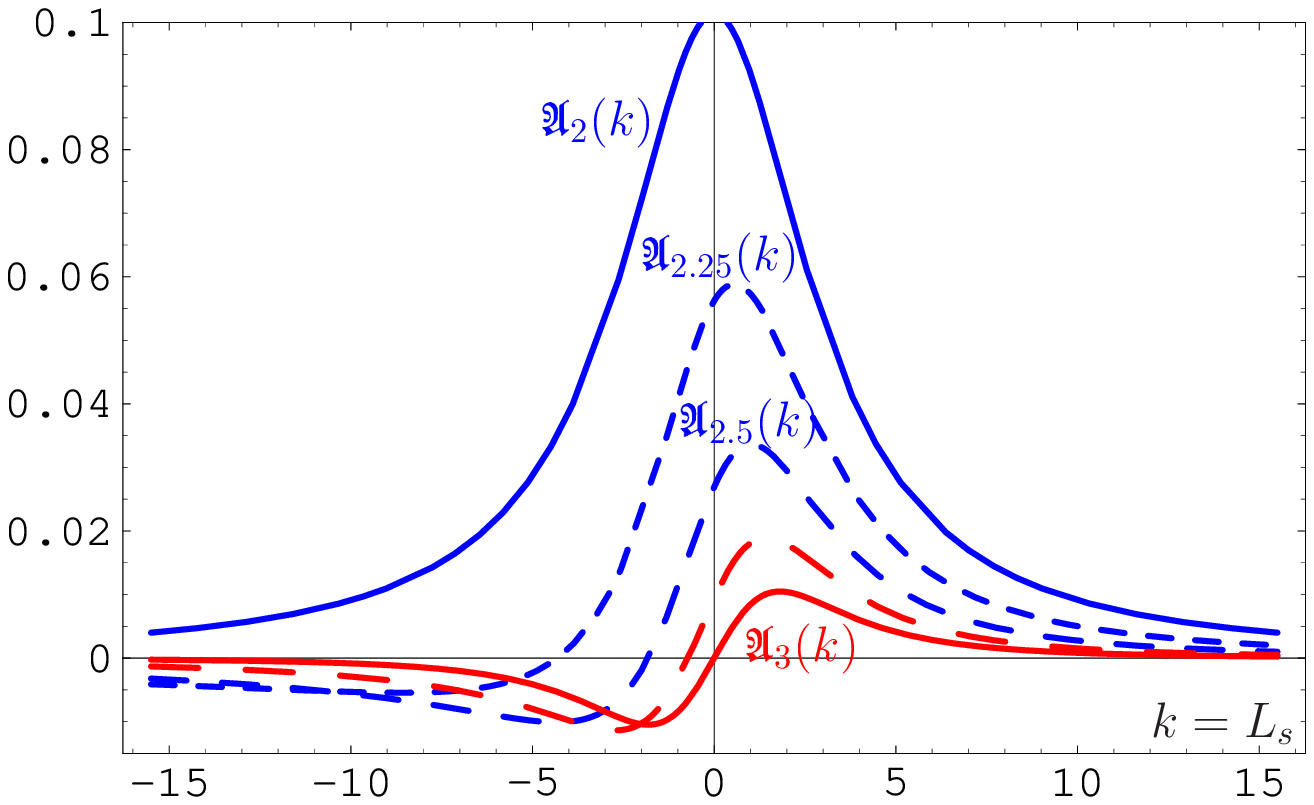}
          ~~~\includegraphics[width=0.45\textwidth]{%fig_anu23_fapt-b.eps
 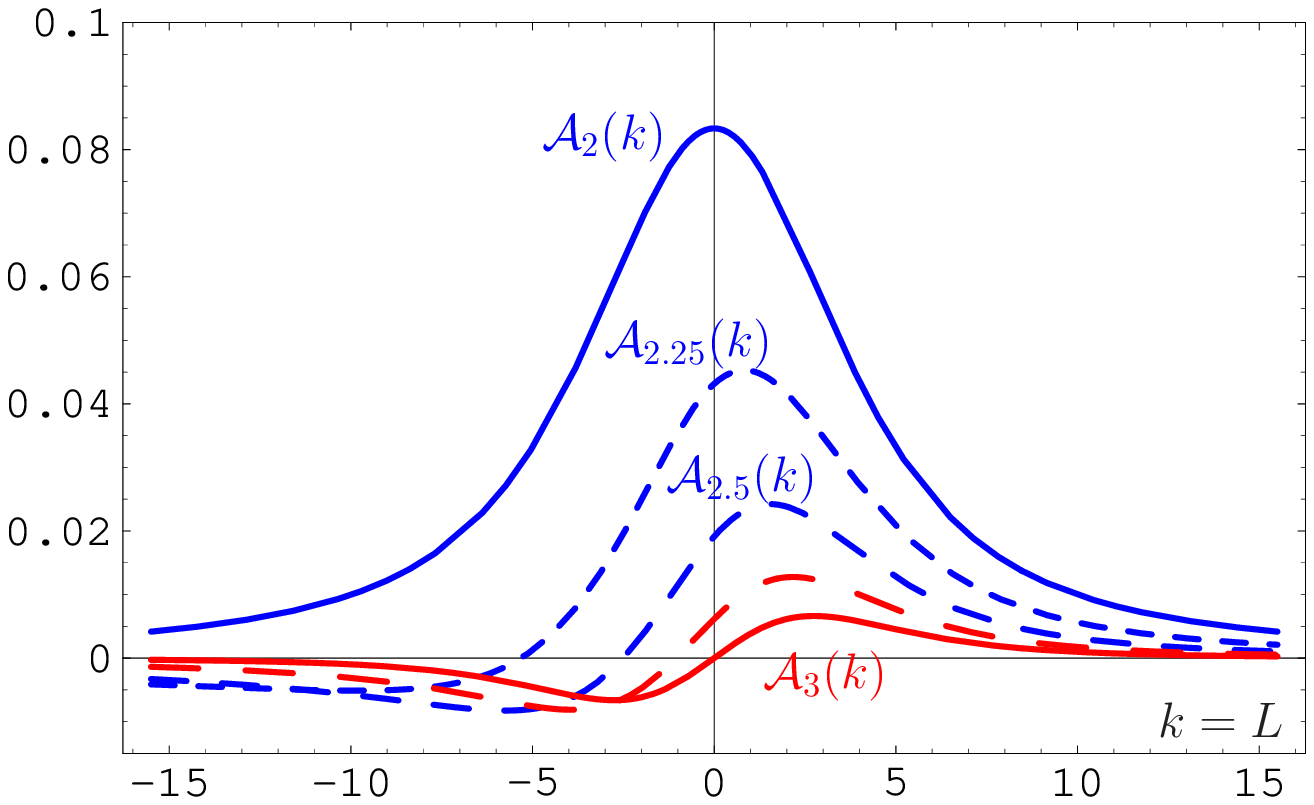}}
   \vspace{-4mm}
   \caption{Comparison of the Minkowski (left panel) and the Euclidean
   (right panel) analytic couplings,
   ${\mathfrak A}_{\nu}(k=L_s)$ and ${\cal A}_{\nu}(k=L)$,
   respectively, for incremental changes of the index $\nu$ in the 
   range 2 to 3.
\label{fig:minkU2-3}}
\end{figure}
%%%%%%%%%%%%%%%%%%%%%%%%%%%%%%%%%%%%%%%%%%%%%%%%%%%%%%%%%%%%%%%%%%%%%%%
\textbf{3.} Moreover, the analytic couplings in both regions $(Q^2,s)$, 
or equivalently, $(L,~L_s)$, have the following symmetry (respectively, 
asymptotic) properties:
\begin{eqnarray}
 \label{eq:U-A-sym}
  \begin{array}{lll}
   {\mathfrak A}_{m}(L_{s}) = (-1)^{m}{\mathfrak A}_{m}(-L_{s})\,,
    & {\cal A}^{}_{m}(L) = (-1)^{m}{\cal A}^{}_{m}(-L)
     & ~\text{for}~ m\geq 2\,,\ m\in \mathbb{N}\,; \\
   {\mathfrak A}_{m}(-\infty) = {\cal A}_{m}(-\infty) = \delta_{m,1}\,,
    & {\mathfrak A}_{m}(\infty) = {\cal A}_{m}(\infty) = 0
     &~\text{for}~ m\in \mathbb{N}\,.
  \end{array}
\end{eqnarray}
%Eq (3.19)
To reveal the details of behavior of the generalized analytic couplings 
${\mathfrak A}_{\nu}(L_s)$ and ${\cal A}_{\nu}(L)$
and make the above statements more transparent, we illustrate them in 
Fig.\ \ref{fig:minkU2-3} in terms of two graphics, which display the 
rate of change of these functions with respect to the index $\nu$ and 
the argument $L$.
Inspection of this figure in conjunction with Eq.\ (\ref{eq:U-A-sym})
provides also information about how the zeros of the couplings occur.

%%%%%%%%%%%%%%%%%%%%%%%%%%%%%%%%%%%%%%%%%%%%%%%%%%%%%%%%%%%%%%%%%%%%%%%
\subsection{Extension to higher loops and convergence of FAPT in the 
            Minkowski vs. the Euclidean region}
\label{subsec:convergence}
%%%%%%%%%%%%%%%%%%%%%%%%%%%%%%%%%%%%%%%%%%%%%%%%%%%%%%%%%%%%%%%%%%%%%%%

The last topic of this section is to extend our results to higher 
loops and to discuss the convergence properties of FAPT in the timelike 
region.
In the following exposition, we shall employ, for the sake of 
simplicity, a special notation for the derivatives with respect to the 
index $\nu\in\mathbb{R}$ of the non-power expansion 
and define
\begin{equation}
 {\cal D}^{k}\,
  {{\cal A}_{\nu}^{} \choose {\mathfrak A}_{\nu}^{}}
   \equiv
    \frac{d^k}{d \nu^k}\,
    {{\cal A}_{\nu}^{} \choose {\mathfrak A}_{\nu}^{}}\,.
 \label{eq:An.Deriv}
\end{equation}
%Eq (3.20)
In our previous paper \cite{BMS05} we have obtained an expansion of the
two-loop analytic coupling ${\cal A }_{\nu}^{(2)}(L)$ in terms of the
one-loop analytic coupling ${\cal A }_{\nu}^{(1)}(L)$ (see Eq.\ (3.29)
in~\cite{BMS05}).
Due to the linearity of this expression in ${\cal A }_{\nu}^{(1)}(L)$,
we can immediately rewrite it for timelike couplings---see Appendix
\ref{app:AnB}, Eqs.\ (\ref{eq:image.u.2.nu}),
(\ref{eq:image.u.3.nu})---by virtue of Eq.\ (\ref{eq:log(a)}) to obtain
the two-loop result
\begin{eqnarray}
 \label{eq:image.u.1}
  {\mathfrak A }_{1}^{(2)}
  = {\mathfrak A }_{1}^{(1)}
  + c_1\nu{\cal D}{\mathfrak A }_{\nu=2}^{(1)}
 &+& c_1^2 \left[{\cal D}^2
                    + {\cal D}
                    - 1
               \right]{\mathfrak A}_{\nu=3}^{(1)}
                    \nonumber \\
 &+& c_1^3 \left[{\cal D}^3
                    + \frac5{2}{\cal D}^2
                    - 2 {\cal D}
                    - \frac1{2}
             \right]{\mathfrak A}_{\nu=4}^{(1)}
 +{\cal O}\left({\cal D}^{4}{\mathfrak A}_{\nu=5}^{(1)}\right)\,,~~~
\end{eqnarray}
%Eq (3.21)
where we employed the auxiliary expansion parameter 
$c_1=b_1/b_{0}^{2}$.

Next, we test the quality of the two-loop expansion of the
analytic-coupling images of $a_{(2)}$ and $\left(a_{(2)}\right)^{2}$ 
in the Minkowski region in comparison with the Euclidean one 
(refraining from displaying the latter because it is completely 
analogous---see Appendix \ref{app:AnB}).
In doing so, we define the following quantities for $\nu=1$ and 
$\nu=2$:
\begin{itemize}
\item
NNLO, i.e., retaining terms up to order $c_1^2$
\begin{eqnarray}
 \label{eq:Delta.MFAPT.2}
  \Delta_3\left({\mathfrak A}_{1}\right)
   &=& \frac{{\mathfrak A}_{1}^{(1)}
          + c_1\,{\cal D}\,{\mathfrak A }_{\nu=2}^{(1)}
          + c_1^2\,\left({\cal D}^{2}+{\cal D}-1\right)
            \,{\mathfrak A }_{\nu=3}^{(1)}}
            {{\mathfrak A}_1^{(2)}}
   \,-\,1\, ;
\end{eqnarray}
%Eq (3.22)
\item
 N$^3$LO, i.e., retaining terms up to order $c_1^3$
\begin{eqnarray}
 \label{eq:Delta.MFAPT.3}
  \Delta_4\left({\mathfrak A}_{1}\right)
   &=& \Delta_3\left({\mathfrak A}_{1}\right)
    + \frac{c_1^3
            \left({\cal D}^3
                 + \frac5{2}\,{\cal D}^2\,
                 - 2\,{\cal D}\,
                 - \frac1{2}
            \right){\mathfrak A}_{\nu=4}^{(1)}}
          {{\mathfrak A}_1^{(2)}}\,;
%Eq (3.23)
\end{eqnarray}
\item N$^4$LO, i.e., retaining terms up to order $c_1^4$
(cf.\ Eq.\ (\ref{eq:u.4.nu}))
\begin{eqnarray}
  {\mathfrak A}_{2}^{(2);\text{MFAPT}}
   &=& {\mathfrak A}_{2}^{(1)}
      + 2\,c_1\,{\cal D}\,{\mathfrak A }_{\nu=3}^{(1)}
      + c_1^2\,\left(3\,{\cal D}^{2}+2\,{\cal D}-2\right)\,
        {\mathfrak A }_{\nu=4}^{(1)}
           \nonumber \\
   & & ~~~~~
      + c_1^3\,\left(4\,{\cal D}^3
                   + 7\,{\cal D}^2\,
                   - 6\,{\cal D}\,
                   - 1
                \right)\,
        {\mathfrak A}_{\nu=5}^{(1)}
   \nonumber \\ \label{eq:MFAPT.A2.3}
   & & ~~~~~
         + c_1^4
            \left(5{\cal D}^4
                + \frac{47}{3}{\cal D}^3
                - 8{\cal D}^2
                - 11{\cal D}
                + \frac{10}{3}
            \right){\mathfrak A}_{\nu=6}^{(1)}~~~
  \label{eq:MFAPT.A2.4}
\end{eqnarray}
%Eq (3.24)
\end{itemize}
with analogous expressions
for $\Delta_3\left({\cal A }_\nu\right)$
and $\Delta_4\left({\cal A }_\nu\right)$,
obtained by using the evident substitution 
${\mathfrak A}\rightarrow {\cal A}$.
%%%%%%%%%%%%%%%%%%%%%%%%%%%%%% F I G U R E  5 %%%%%%%%%%%%%%%%%%%%%%%%%
\begin{figure}[ht]
 \centerline{\includegraphics[width=0.47\textwidth]{%fig-fapt_2-3_lamb_it2.eps
   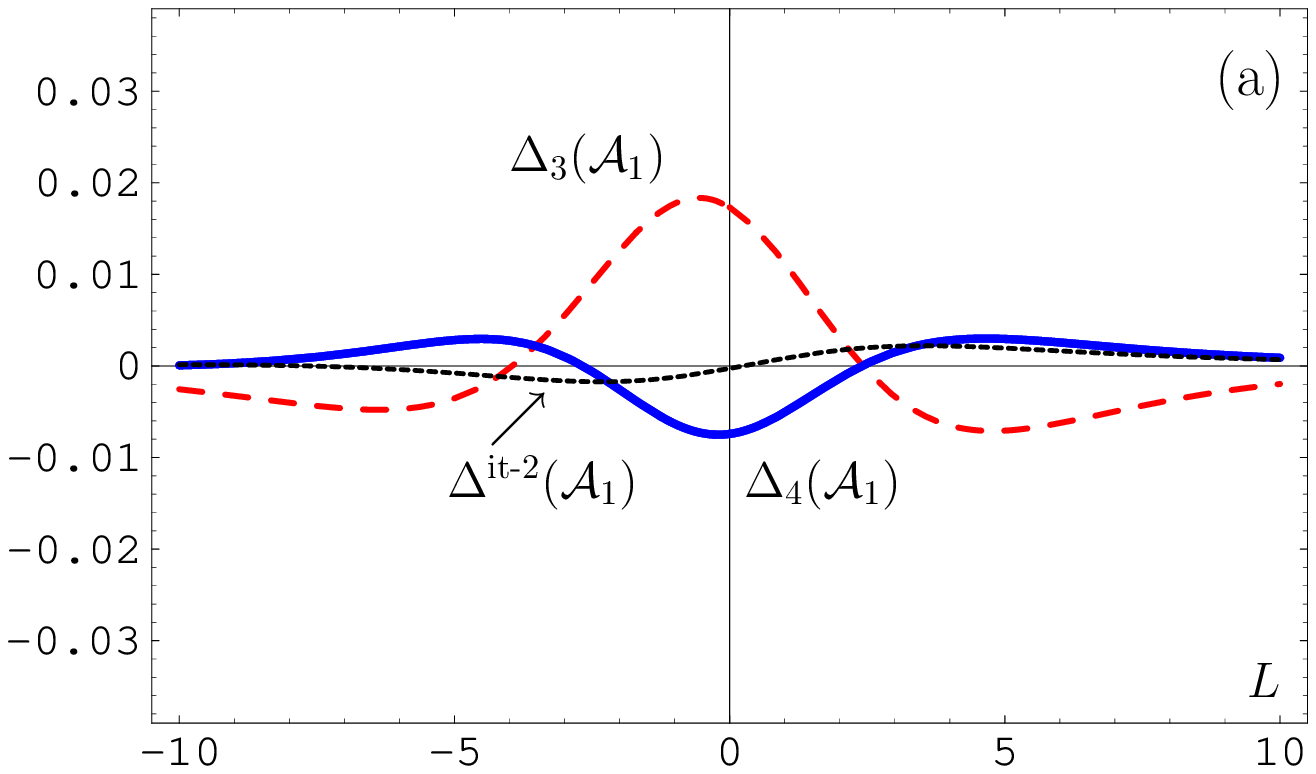}~~~%
   \includegraphics[width=0.47\textwidth]{%fig-mfapt_2-3_lamb_it2.eps
    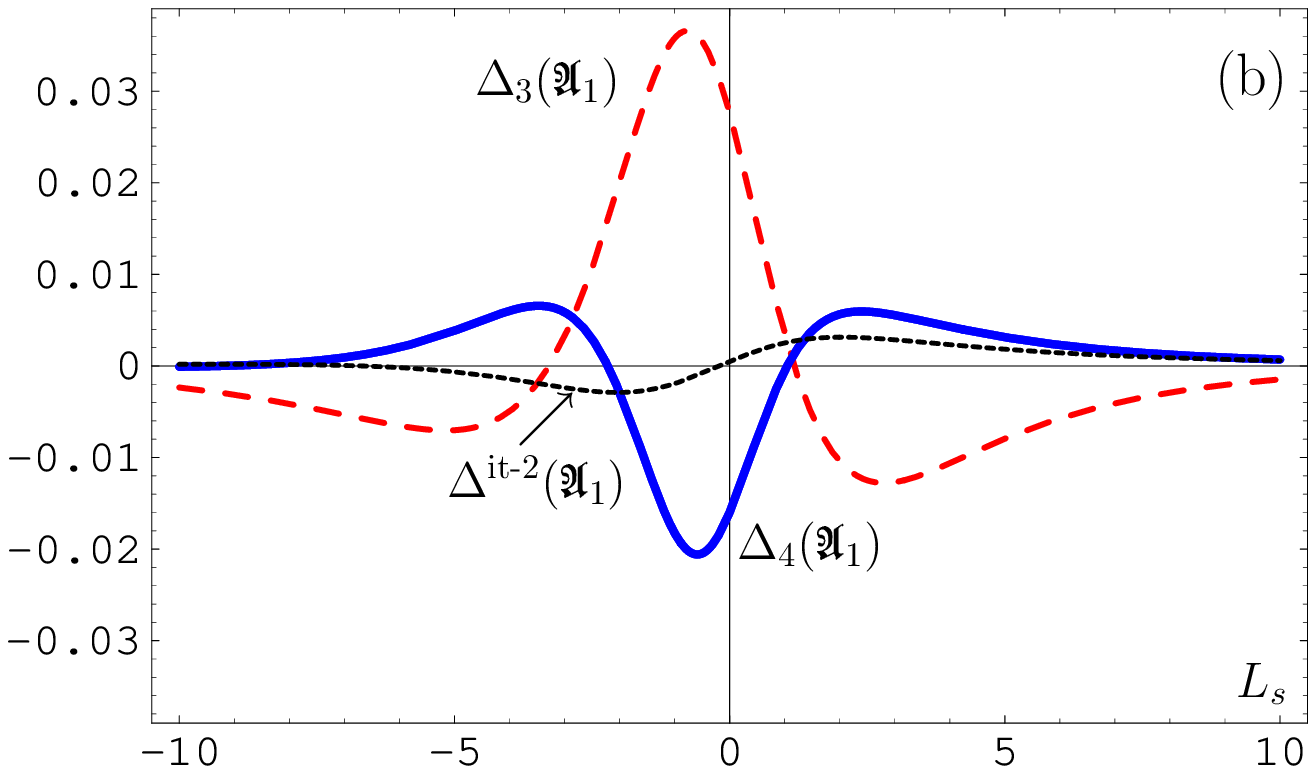}}
   \caption{\label{fig:FAPT-MFAPT-1}\footnotesize
    (a): The dashed red line corresponds to $\Delta_{3}({\cal A}_{1})$,
    whereas the solid blue line represents $\Delta_{4}({\cal A}_{1})$.
    (b): The dashed red line denotes $\Delta_{3}({\mathfrak A}_{1})$
    and the solid blue line $\Delta_{4}({\mathfrak A}_{1})$.
    The dotted black line indicates $\Delta^\text{it-2}({\cal A}_1)$
    (left panel) and
    $\Delta^\text{it-2}({\mathfrak A}_1)$ (right panel).}
\end{figure}
%%%%%%%%%%%%%%%%%%%%%%%%%%%%%%%%%%%%%%%%%%%%%%%%%%%%%%%%%%%%%%%%%%%%%%%

Figure \ref{fig:FAPT-MFAPT-1} illustrates the convergence quality of 
the FAPT expansion in the spacelike and the timelike regions in terms 
of the quantities $\Delta_{n}$, defined above, that encapsulate the 
deviation of our approximations from the exact results.
In addition, we consider the auxiliary quantity $\Delta^{\text{it}-2}$
(dotted black line) for the spacelike ($L$) and the timelike ($L_s$) 
regions, defined by the respective expressions
\begin{equation}
\Delta^{\text{it}-2}({\cal A}_{\nu}^{})
=
  \frac{{\cal A}_{\nu}^{(2)\text{it}-2}-
  {\cal A}_{\nu}^{(2)}}{{\cal A}_{\nu}^{(2)}} \, ; \quad
  \Delta^{\text{it}-2}({\mathfrak A}_{\nu})
=
  \frac{{\mathfrak A}_{\nu}^{(2)\text{it}-2}-
  {\mathfrak A}_{\nu}^{(2)}}{{\mathfrak A}_{\nu}^{(2)}}\, ,
\label{eq:delta-it}
\end{equation}
%Eq (3.25)
which, as one sees, result from replacing the exact spectral density
(cf.\ (\ref{eq:Spec.Den.Lamb})) by its second iteration (details are
relegated to Appendix \ref{app:spectr}).

From Fig.\ \ref{fig:FAPT-MFAPT-1} one observes that the convergence of
${\cal A }_{1}^{(2)}$ and ${\mathfrak A}_{1}^{(2)}$ at the N$^3$LO of 
the expansion in the auxiliary parameter $c_1$ is sufficiently 
accurate.
In the Euclidean case (left panel), the errors---defined by
$\Delta_3^\text{FAPT}\left({\cal A }_1\right)$,
(dashed red line) and
$\Delta_4^\text{FAPT}\left({\cal A }_1\right)$
(solid blue line)---induced by the non-power-series expansion are by a 
factor of two less than those in the Minkowski case (right panel): 
$\Delta_3^\text{MFAPT}\left({\mathfrak A}_{1}\right)$
(dashed red line) and 
$\Delta_4^\text{MFAPT}\left({\mathfrak A}_{1}\right)$ 
(solid blue line).
The largest error results around $L=L_s=0$, but already for $|L|>2$ the
uncertainty is less than a few per mil, rendering the convergence of
the expansion highly reliable.
Deriving the two-loop explicit expression 
\begin{subequations}
 \begin{eqnarray}
  \label{eq:u.2.nu}
  {\mathfrak A}_{\nu}^{(2)}
   &=& {\mathfrak A}_{\nu}^{(1)}
     + c_1\,\nu\,{\cal D}\,
       {\mathfrak A}_{\nu+1}^{(1)}
     + c_1^2\,\nu\left[\frac{\nu+1}{2!}\,{\cal D}^2\,
                     + \,{\cal D}\,
                     - \,1
                 \right]
       {\mathfrak A}_{\nu+2}^{(1)}~~~\\
  \label{eq:u.3.nu}
   &+& c_1^3\,\nu\left[\frac{(\nu+1)(\nu+2)}{3!}\,{\cal D}^3\,
                     + \,\frac{2\,\nu+3}{2}\,{\cal D}^2
                     + \,(1 +\nu ){\cal D}\,
                     - \,\frac12
                 \right]
       {\mathfrak A}_{\nu+3}^{(1)}~~~\\
   &+& c_1^4\,\nu\left[\frac{(\nu+1)(\nu+2)(\nu+3)}{4!}\,{\cal D}^4\,
                     + \frac{3\,\nu^2+12\,\nu+11}{6}\,{\cal D}^3
                     - \frac{\nu^2+2\,\nu}{2}\,{\cal D}^2 \right.
   \nonumber\\
             &&\left.  ~~~~~~~         - \frac{3\,\nu+5}{2}\,{\cal D}\,
                     + \frac{3\,\nu+4}{6}
                  \right]
       {\mathfrak A}_{\nu+4}^{(1)}
  + {\cal O}\left({\cal D}^{5}\,{\mathfrak A}_{\nu+5}^{(1)}
            \right)\, ,
\label{eq:u.4.nu}
\end{eqnarray}
\end{subequations}
%Eqs (3.26a) (3.26b) (3.26c)
that is extended to the three-loop order of the running coupling in 
Appendix \ref{app:AnB} (Eqs.\ (\ref{eq:image.u.2.nu}) 
to (\ref{eq:image.u.4.nu})), 
a similar quality of convergence can be established for any desired 
fractional value of the index $\nu$.

%%%%%%%%%%%%%%%%%%%%%%%%%%%%%%% F I G U R E  6 %%%%%%%%%%%%%%%%%%%%%%%%%
\begin{figure}[th]
 \centerline{\includegraphics[width=0.47\textwidth]{%fig-a2_4-mag-lamb.eps
   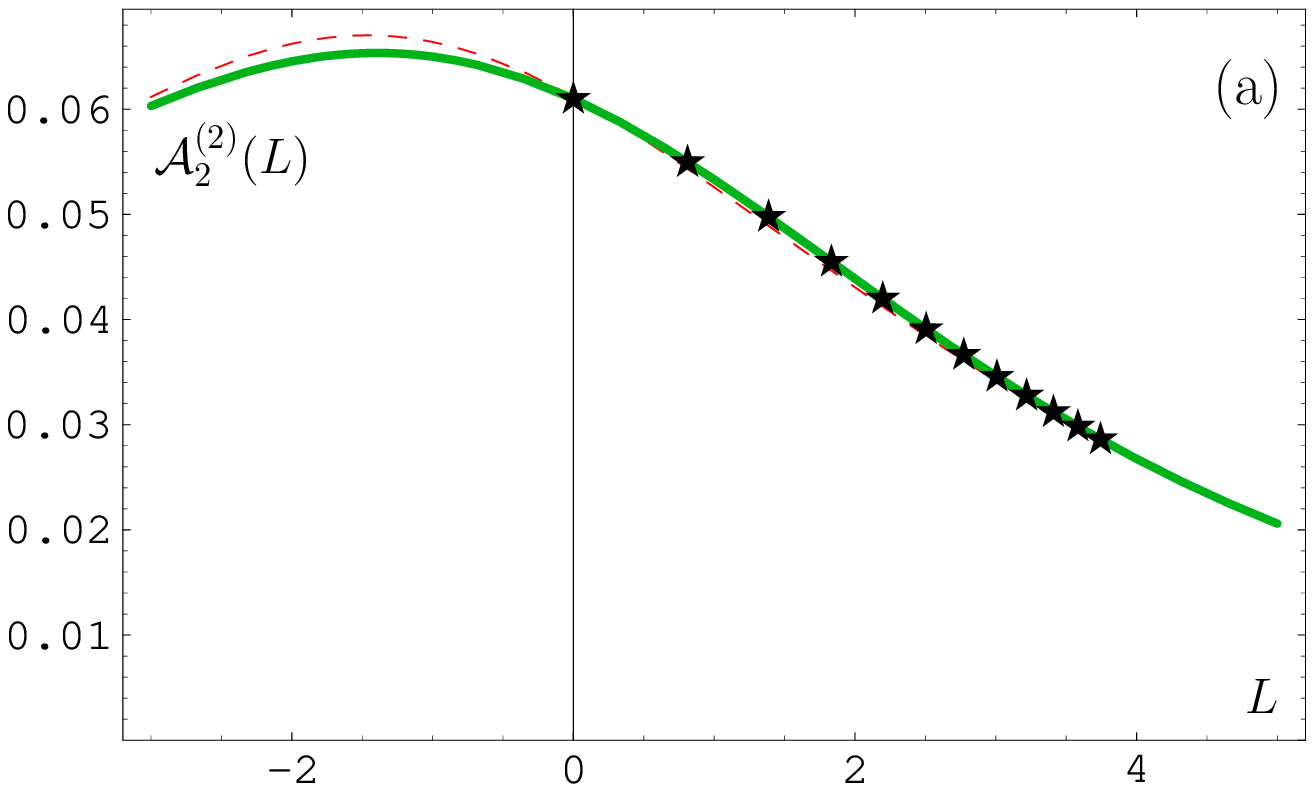}~~~%
   \includegraphics[width=0.47\textwidth]{%fig-u2_5-mag-lamb.eps
    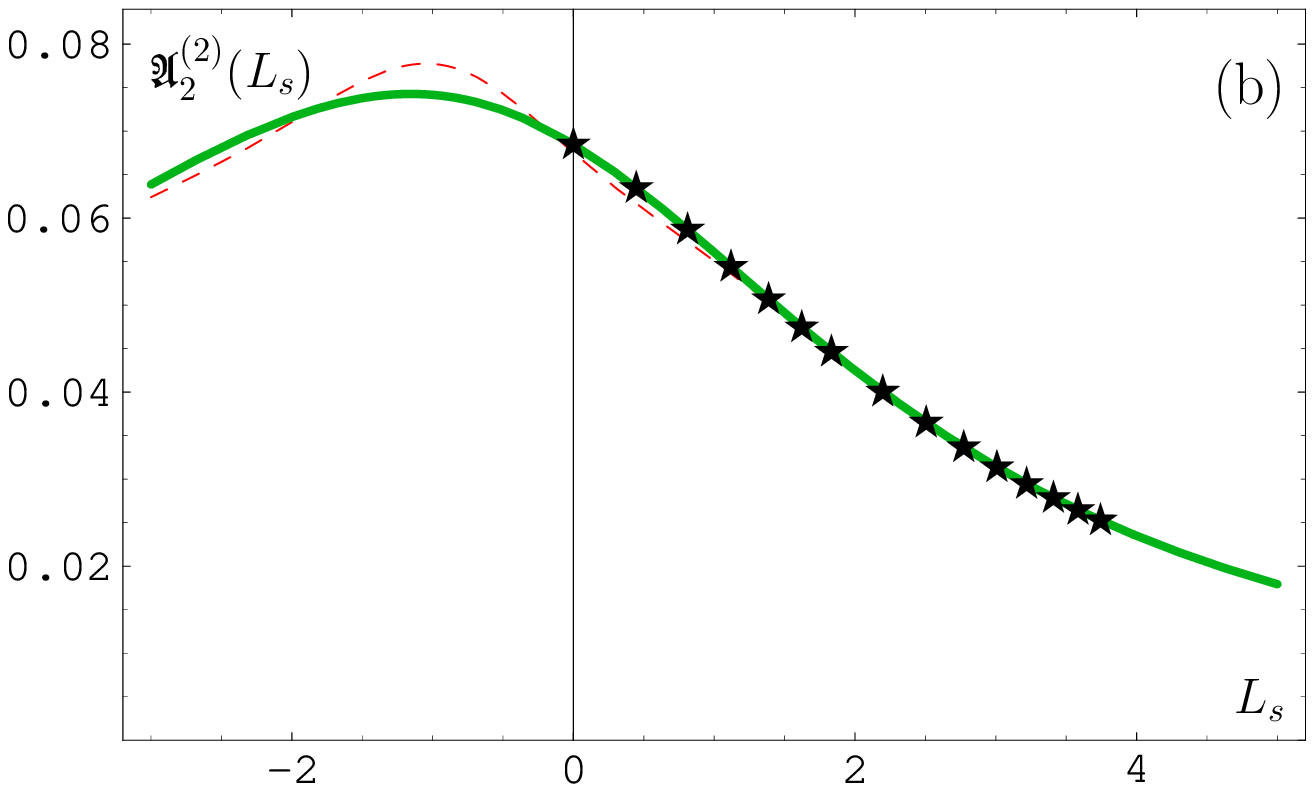}}
   \caption{\label{fig:FAPT-MFAPT-2}\footnotesize
    (a): The dashed red line corresponds to
    ${\cal A}_2^{(2);\text{FAPT}}(L)$, computed analytically via 
    Eqs.\ (\ref{eq:image.u.2.nu})--(\ref{eq:image.u.4.nu}) 
    (for $c_2=0$), whereas the solid green line represents
    the exact expression for ${\cal A}_2^{(2)}(L)$, cf.\ Eqs.\
    (\ref{eq:A.E}) and (\ref{eq:Spec.Den.Lamb}).
    (b): The dashed red line denotes
    ${\mathfrak A}_2^{(2);\text{MFAPT}}(L_s)$, computed analytically
    via Eq.\ (\ref{eq:MFAPT.A2.3}), whereas the solid green line
    represents the exact result for ${\mathfrak A}_2^{(2)}(L_s)$,
    given by Eq.\ (\ref{eq:Mink.U2.2Loop}).
    On both panels, we also indicate by \ding{72} the results of
    Magradze, presented in \cite{KM03,Mag03u}.
    }
\end{figure}
%%%%%%%%%%%%%%%%%%%%%%%%%%%%%%%%%%%%%%%%%%%%%%%%%%%%%%%%%%%%%%%%%%%%%%%
In  Figure \ref{fig:FAPT-MFAPT-2} we show the results of the exact and
the approximate calculation of ${\cal A}_2^{(2);\text{FAPT}}(L)$ and
${\mathfrak A}_2^{(2);\text{MFAPT}}(L_s)$.
This figure also includes the results (denoted by the symbol \ding{72})
obtained before by Magradze \cite{Mag03u}.
One observes that the agreement between his numerical estimates and our
more elaborated calculations is excellent.
We take the opportunity to remark that our opposite statements in
\cite{BMS05}, notably Figure 4, were incorrect owing to an error in
our code.
This bug has now been eliminated, so that Magradze's numerical results
are fully supported by our calculation of the exact expressions
for both analytic images ${\cal A }_{2}^{(2)}$ and
${\mathfrak A}_{2}^{(2)}$.

In concluding this section, it is worth listing the advantages of our 
calculation:
(i)   a good convergence of the non-power-series expansion,
(ii)  full control over the region in $L_s$ (correspondingly $L$), 
      in which the maximum uncertainty occurs, 
      making possible systematic improvements, and
(iii) and most important for practical applications, a considerably
      reduced uncertainty level of the expansion in the physically
      interesting region, say, beyond $1$~GeV, (which corresponds to 
      $L=2$) of only a few per mil.

%%%%%%%%%%%%%%%%%%%%%%%%%%%%%%%%%%%%%%%%%%%%%%%%%%%%%%%%%%%%%%%%%%%%%%%
%%%%%%%%%%%%%%%%%%%%%%%%%%%%%%%%%%%%%%%%%%%%%%%%%%%%%%%%%%%%%%%%%%%%%%%
\section{Scalar correlator and Higgs boson decay into hadrons in MFAPT}
\label{sec:higgs}
%%%%%%%%%%%%%%%%%%%%%%%%%%%%%%%%%%%%%%%%%%%%%%%%%%%%%%%%%%%%%%%%%%%%%%%

To bring out the contrast with the original APT and effect the
advantages and the utility of the (M)FAPT machinery with regard to the 
conventional perturbative expansion, we consider in this section the 
decay width of the Higgs boson into a bottom-antibottom pair, 
$\Gamma(\text{H} \to b\bar{b})$, using a multi-loop approximation. 
We stress in this context again that, though there are no ghost 
singularities in the Minkowski space owing to the strong coupling, 
the analytic continuation from the spacelike to the timelike region
entails so-called `kinematical' $\pi^2$ terms (see, for instance, 
\cite{BCK05}) that can amount to pretty large 
contributions as the order of the perturbative expansion increases.
Hence, even at high energies, relevant for the Higgs decay, the 
inclusion of these terms to all orders of the perturbative expansion is 
mandatory, albeit a difficult task.
It is exactly this issue that singles out the utility of MFAPT because 
within such a perturbative approach, the analytic couplings contain 
all the aforementioned $\pi^2$ terms inherently by construction.

Our calculation of $\Gamma(\text{H} \to b\bar{b})$ via $R_\text{S}$ 
will be carried out in the \MS scheme and will include the evolution 
of both the coupling and the $b$-quark mass.
Special attention will be given below to the origin of the $\alpha_{s}$
corrections in order to distinguish between the loop expansion and the 
loop evolution. 
In the following, we will compare our results with those obtained in 
the \MS scheme in Refs.\ \cite{Che96,ChKS97,BKM01,BCK05}, where the 
notation $\Ds a_{s}=\alpha_{s}/\pi$ (called ``couplant'') was 
extensively used.
For the sake of a better presentation of our results in comparison with 
the existing ones, just mentioned, it is useful to introduce the 
following abbreviation 
\begin{eqnarray}
 \label{eq:Che.Anal.Coupl}
  \left[\left(a_s(s)\right)^\nu\right]_\text{an} =
  {\mathfrak a}_\nu^{(l)}(s)
   \equiv \left(\frac{4}{b_0}\right)^{\nu}\,
           {\mathfrak A}_{\nu}^{(l)}(s)\,.
\end{eqnarray}
%Eq (4.1)

%%%%%%%%%%%%%%%%%%%%%%%%%%%%%%%%%%%%%%%%%%%%%%%%%%%%%%%%%%%%%%%%%%%%%%%
\subsection{Standard perturbation-theory analysis of 
            $\mathbf{R}_\text{S}$}
\label{subsec:stand-Higgs}                                                                
%%%%%%%%%%%%%%%%%%%%%%%%%%%%%%%%%%%%%%%%%%%%%%%%%%%%%%%%%%%%%%%%%%%%%%%

The Higgs-boson decay into a bottom-antibottom pair can be expressed 
in QCD by means of the correlator
$$\Pi(Q^2)
= (4\pi)^2 i\int dx e^{iqx}\langle 0|\;T[\;J^\text{S}_b(x)
J^\text{S}_{b}(0)\,]\;|0\rangle$$
of two quark scalar (S) currents in terms of the discontinuity
of its imaginary part \cite{Djo05}, i.e.,
$
  R_\text{S}(s)
 =
  \textbf{Im}\, \Pi(-s-i\epsilon)/{(2\pi\,  s)}
$,
so that the width reads
\begin{eqnarray}
\Gamma(\text{H} \to b\bar{b})
=\frac{G_F}{4\sqrt{2}\pi}M_\text{H}
m_{b}^{2}(M_\text{H}) R_\text{S}(s = M_\text{H}^2)
\label{decay_rate_for_b}\, .
\end{eqnarray}
%Eq (4.2)
Above, $Q^2 = - q^2$ and $J^\text{S}_b=\bar{\Psi}_b\Psi_b$ is the
scalar current for bottom quarks with mass $m_b$, coupled to the
scalar Higgs boson with mass $M_\text{H}$.
Direct multi-loop calculations are usually performed in the Euclidean
(spacelike) region for the corresponding Adler function $D_\text{S}$   
\cite{Che96,BCK05,BKM01}, where QCD perturbation theory works.
Hence, we write                  
\begin{eqnarray}
 \widetilde{D}_{\text{S}}(Q^2;\mu^2)
  &=& 3\,m_b^2(Q^2)
       \left[1+\sum_{n \geq 1} d_n(Q^2/\mu^2)~a_{s}^n(\mu^2)
        \right]\,,
\label{eq:D-s}
\end{eqnarray}
%Eq (4.3)
using, as announced above, the notation $\Ds a_{s}={\alpha_{s}/ \pi}$.
Connecting to our discussion in Sec.\ \ref{sec:basics} below
Eq.\ (\ref{eq:R}), and adjusting to the scalar case, we now write the  
results of the `standard machinery' (see, for instance, \cite{ChKS97}):
\begin{eqnarray}
  \widetilde{R}_\text{S}(s)
\equiv
  \widetilde{R}_\text{S}(s,s)
=
  3m^{2}_{b}(s)\left[ 1 + \sum_{n\geq 1}^{} r_{n}~a_{s}^n(s)
               \right]\, .
  \label{eq:R-s}
\end{eqnarray}
%Eq (4.4)
The coefficients $r_n$ contain characteristic `$\pi^2$ terms' due to
the (integral) transformation $\hat{R}$ of the powers of the
logarithms appearing in $\widetilde{D}_{\text{S}}$ 
(cf.\ Eq.\ (\ref{eq:D-s})), as it was discussed in the beginning of 
Sec.\ \ref{sec:basics}.
Recall that in the case of $\widetilde{R}_\text{S}(s)$, these 
logarithms stem from two different sources: one is the running of 
$\alpha_{s}$ in $\widetilde{D}_\text{S}$,
in analogy to Eq.\ (\ref{eq:R}), the other is related to the evolution 
of $m_{b}^{2}(Q^2)$.
Therefore, the coefficients $r_n$ in (\ref{eq:R-s}) appear to 
be related to a) the coefficients $d_n$ in (\ref{eq:D}), the latter 
being directly calculable in the Euclidean space, and b) to a 
combination of the mass anomalous dimension $\gamma_i$ and the 
$\beta$-function coefficients $b_j$, multiplied by `$\pi^2$ powers' 
\cite{BKM01,Che96,ChKS97}.
It turns out that the influence of these $\pi^2$ terms can be
substantial, as the following quite recent result, derived in
\cite{BCK05}, demonstrates:
\begin{eqnarray}
 \left[{3m_b^2}\right]^{-1}\widetilde{R}_\text{S}
 &=& 1
   + 5.667\,a_s
   + a_s^2\left[51.57
             - \unl{15.63}
             - N_f\left({1.907} - \unl{0.548}\right)
          \right]
 \nonumber\\
 && ~\,
   + a_s^3\left[648.7 - \unl{484.6}
             - N_f\left(63.74 - \unl{37.97}\right)
             + N_f^2\left(0.929 - \unl{0.67}\right)
          \right]
 \nonumber\\
 && ~\,
   + a_s^4\left[9470.8 - \unl{9431.4}
             - N_f\left(1454.3 - \unl{1233.4}\right)
          \right.
 \nonumber\\
 && ~~~~~~~~~~
          \left.
             + N_f^2\left(54.78 - \unl{45.10}\right)
             - N_f^3\left(0.454 -\unl{0.433}\right)
          \right]\,.
 \label{eq:RSnumx}
\end{eqnarray}
%Eq (4.5)
For emphasis, we have underlined the contributions of those 
$\pi^2$ terms which originate from the analytic continuation.
As one can readily verify, the total amount of these terms is of the
order of the original coefficients $d_n$, in particular, as regards 
the coefficient $d_4$.
This makes it apparent that such terms have to be taken into account
in all orders of the perturbative expansion.
We stress that this is exactly the advantage provided by the 
analytic machinery, developed here and in \cite{BMS05}, and this 
conceptual advantage arises naturally without any additional 
optimization procedure. 
Indeed, in FAPT we do not need to expand the renormalization factors
into a truncated series of logarithms; instead we can transform 
them `as a whole' by means of the $\textbf{A}_\text{M}$-operation.

To complete the presentation of the standard analysis, let us display
the final result at the ${\cal O}(a_s^4)$, 
taken from Ref.\ \cite{BCK05}:
\begin{eqnarray}
 \left[{3m_b^2}\right]^{-1}\widetilde{R}_\text{S}
 &=& 1 + 5.6668\,a_s
       + 29.147\,a_s^2
       + 41.758\,a_s^3
       - 825.7\,a_s^4
       \label{eq:R-Ch}      \\
 &=& 1 + 0.2075
       + 0.0391
       + 0.0020
       - 0.00148\,.
 \label{eq:R-Ch-numer}
\end{eqnarray}
%Eqs (4.6) (4.7)
Note that in Eq.\ (\ref{eq:R-Ch-numer})
$a_{s}=a_{s}(M_\text{H}^2)= 0.0366$
is chosen, which corresponds to the Higgs boson mass
$M_\text{H} = 120$~GeV.

%%%%%%%%%%%%%%%%%%%%%%%%%%%%%%%%%%%%%%%%%%%%%%%%%%%%%%%%%%%%%%%%%%%%%%%
\subsection{FAPT analysis of $\widetilde{R}_\text{S}$}
\label{subsec:FAPT-R}
%%%%%%%%%%%%%%%%%%%%%%%%%%%%%%%%%%%%%%%%%%%%%%%%%%%%%%%%%%%%%%%%%%%%%%%

We turn now our attention to effects related to the renormalization of 
the bottom-quark mass.
For the running mass $m_{(l)}(Q^2)$, in the $l$-loop approximation, 
one has the following general solution of the RG equation
\begin{eqnarray}
\label{eq:m2-def}
  m_{(l)}^2(Q^2)
  &=& m_{(l)}^2(\mu^2)\,
       \exp\left[2\int_{a_{s}(\mu^2)/4}^{a_{s}(Q^2)/4}
                 \frac{\gamma_{m}(x)}{\beta(x)}\, dx
                            \right] \\
  &=& m_{(l)}^2(\mu^2)\,
       \frac{\left[a_{s}(Q^2)\right]^{\nu_0}
              f_{(l)}(a_s(Q^2))}
            {\left[a_{s}(\mu^2)\right]^{\nu_0}
              f_{(l)}(a_s(\mu^2))}\,,
\label{eq:m-run}
\end{eqnarray}
%Eqs (4.8) (4.9)
where
\begin{eqnarray}
 \nu_0 = 2\frac{\,\gamma_0}{b_0}
 \label{eq:nu.0.nu.1}
\end{eqnarray}
%Eq (4.10)
and the function $f_{(l)}(a_s)$, given by
\begin{eqnarray}
 f_{(l)}(a_s) 
 = \exp\left[2\int_{0}^{a_{s}/4}
             \left(\frac{\gamma_{m}^{(l)}(x)}{\beta^{(l)}(x)}
                  -\frac{\gamma_0\,x
                         }{b_0\,x^2 
                         }\right)\, dx
                   \right]\,,
\label{eq:phi}
\end{eqnarray}
%Eq (4.11)
accumulates the effects of the second- and higher-loop evolution of
$m_{(l)}^2(Q^2)$ with $Q^2$.
In the one-loop approximation ($l=1$), $f_{(l)}(a_{s})$ is set by 
definition equal to unity.
On the other hand, for $l=2$ and $l=3$ we obtain
\begin{eqnarray}
 f_{(2)}(a_s) 
 = \left[1 + \delta_1\,a_s\right]^{\nu_1}
 \quad\text{with~~~}
 \delta_1 = \frac{b_1}{4 b_0} = \frac{c_1 b_0}{4}\,,~
 \nu_1 = 2 \left(\frac{\gamma_1}{b_1}
                 -\frac{\gamma_0}{b_0}
           \right)
\label{eq:f_2}
\end{eqnarray}
%Eq (4.12) 
and
\begin{subequations}
\begin{eqnarray}
 f_{(3)}(a_s) 
 = \left[1 + \delta_{1}\,a_s+ \delta_{2}\,a_s^2\right]^{\nu_{20}}
    \exp\left[\nu_{21}\,
              \arccos\left(
                   \frac{1+\delta_{1}\,a_s/2}
                        {\sqrt{1 + \delta_{1}\,a_s+ \delta_{2}\,a_s^2}}
                      \right)
        \right]
\label{eq:f_3}
\end{eqnarray}
%Eq (4.13a) 
with
\begin{eqnarray}
 \label{eq:delta_21}
  \delta_2 = \frac{b_2}{16 b_0}\,,~
  \nu_{20} = \left(\frac{\gamma_2}{b_2}
                  -\frac{\gamma_0}{b_0}
             \right)\,,~
  \nu_{21} = \frac{-2\,b_1}
                  {\sqrt{4b_2b_0-b_1^2}} 
              \left(\frac{\gamma_2}{b_2}
                   -2\frac{\gamma_1}{b_1}
                   +\frac{\gamma_0}{b_0}
              \right)\,.
\end{eqnarray}
%Eq (4.13b) 
\end{subequations}
Introducing the RG-invariant quantity $\hat{m}_{(l)}$, see, e.g.,
\cite{BKM01,KKP00},
\begin{eqnarray}
\label{eq:m2-hat}
  \hat{m}_{(l)}
   &=& m_{(l)}(\mu^2)
             \left\{\left[a_{s}(\mu^2)\right]^{\nu_0}
              f_{(l)}(a_s(\mu^2))\right\}^{-1/2}\,,
\end{eqnarray}
%Eq (4.14)
one can rewrite Eq.\ (\ref{eq:m-run}) in the form
\begin{eqnarray}
 \label{eq:m2-hat-run}
 m_{(l)}^2(Q^2)
  &=& \hat{m}_{(l)}^2
      \left[a_{s}(Q^2)\right]^{\nu_0}
      f_{(l)}(a_s(Q^2))\,,
\end{eqnarray} 
%Eq (4.15)     
where the expansion of $f_{(l)}(x)$ at the three-loop order is given by
\begin{eqnarray}
  f_{(l)}(a_s)
   &=& 1
    +  a_s\,\frac{b_1}{2b_0}\left(\frac{\gamma_1}{b_1} 
    - \frac{\gamma_0}{b_0}\right)
    +  a_s^2\,\frac{b_1^2}{16\,b_0^2}
            \left[\frac{\gamma_0}{b_0}-\frac{\gamma_1}{b_1}
                 + 2\,\left(\frac{\gamma_0}{b_0}
                 -\frac{\gamma_1}{b_1}\right)^2
                 + \frac{b_0 b_2}{b_1^2}\left(\frac{\gamma_2}{b_2}
                 -\frac{\gamma_0}{b_0}\right)
           \right]\nonumber\\
   &+& O\left(a_s^3\right)\,,
 \label{varphi}
\end{eqnarray}
%Eq (4.16)
which is in one-to-one correspondence with Eq.\ (15) found by Chetyrkin 
in \cite{Che97}.
[Note, however, that Chetyrkin expands instead the expression 
 $\sqrt{f_{(l)}(x)}\equiv c(x)$
 and uses different normalizations; viz.,
 $\bar{\beta}_n=b_n/(4^nb_0)$ and 
 $\bar{\gamma}_n=\gamma_n/(4^nb_0)$.] 
Keeping in mind the main purpose of our task, namely, the sequential 
``analytization'' of $D$, we rewrite the RG Eq.\ (\ref{eq:m2-hat-run}) 
in the form of a power series to get
\begin{eqnarray}
 \label{eq:Z-2loop}
   m_{(l)}^2(Q^2)
    &=& \hat{m}_{(l)}^2\, 
         \left(a_{s}(Q^2)\right)^{\nu_0}
          \left[1
                + \sum_{m\geq 1}^{\infty}
                   e^{(l)}_m\,
                   \left(a_{s}(Q^2)\right)^m
              \right]\,,
\end{eqnarray}
%Eq (4.17)
where the coefficients $e^{(l)}_m$ depend implicitly on the RG
parameters via Eq.\ (\ref{eq:phi}).
For the simplest case of a two-loop running, they can be written 
down explicitly:
\begin{eqnarray}
 e^{(2)}_m 
 &=& \frac{\Gamma(\nu_1+1)\,}
          {\Gamma(m+1)\Gamma(\nu_1-m+1)}(\delta_1)^m
          \,.
\label{eq:f-term}
\end{eqnarray}
%Eq (4.18)
Substituting Eq.\ (\ref{eq:Z-2loop}) for $m_{(l)}^2(Q^2)$ into
$\widetilde{D}_{\text{S}}(Q^2)\equiv \widetilde{D}_{\text{S}}(Q^2;Q^2)$
in Eq.\ (\ref{eq:D-s}), one finds 
\begin{eqnarray}
 \left[3\,\hat{m}_b^2\right]_{(l)}^{-1}\, 
  \widetilde{D}^{(l)}_{\text{S}}(Q^2)
  = \left(a^{(l)}_{s}(Q^2)\right)^{\nu_0}
    + \sum_{n\geq1}^{l}d_n\,
       \left(a^{(l)}_{s}(Q^2)\right)^{n+\nu_0}
    + \sum_{m\geq1}^{\infty}\Delta^{(l)}_m\,
       \left(a^{(l)}_{s}(Q^2)\right)^{m+\nu_0}    
 \label{eq:D-approx}
\end{eqnarray}
%Eqs (4.19) 
with
\begin{eqnarray}
 \Delta^{(l)}_m
  &=& e_m^{(l)}
   + \sum_{ k\geq1}^{\textbf{min}[l,m-1]}d_k\,e_{m-k}^{(l)}\,.
 \label{eq:d-tild_1}
\end{eqnarray}
%Eq (4.20)
The effects of mass evolution of the higher orders are collected
in the third term on the RHS of Eq.\ (\ref{eq:D-approx}) and have 
been purportedly separated from the original series expansion of 
$D$ (truncated at $n=l$), the latter being represented by the 
second term on the RHS of Eq.\ (\ref{eq:D-approx}).
In practice, for $Q\geq 2$~GeV, i.e., for $\alpha_s\leq0.4$, 
the truncation at $m=l+4$ of the summation  (\ref{eq:Z-2loop}) 
produces a truncation error smaller than $0.01\%$.

To obtain $\widetilde{R}_\text{S}^{\text{MFAPT}}$, we recall the 
action of the $\textbf{A}_\text{M}$ operation in FAPT, as described 
in Sec.\ \ref{sec:basics}, and consider the map of the quantity 
$\widetilde{D}_{\text{S}}^{(l)}(Q^2)$ in Eq.\ (\ref{eq:D-approx}) 
onto the Minkowski region.
Following the ``analytization'' procedure illustrated in 
Fig.\ \ref{fig:APT-scheme}, we then obtain
\begin{eqnarray}
&&  \widetilde{R}_\text{S}^{(l)\text{MFAPT}}
   =  \textbf{A}_\text{M}[D^{(l)}_{\text{S}}]
   =  3\,\hat{m}_{(l)}^2\,
      \left[{\mathfrak a}_{\nu_{0}}^{(l)}
          + \sum_{n\geq1}^{l} d_{n}^{}
            {\mathfrak a}_{n+\nu_{0}}^{(l)}
          + \sum_{m\geq1}^{}\Delta_{m}^{(l)}
            {\mathfrak a}_{m +\nu_{0}}^{(l)}                          
      \right]\,, \label{eq:R-MFAPT}
\end{eqnarray}
%Eq (4.21)
where the superscript $l$ denotes the loop order of the evolution and 
fixes at the same time the order of the perturbative expansion of the 
$D_S$-function.
The above expression contains, by means of the coefficients 
$\Delta_{n}^{(l)}~(~e^{(l)}_k)$
and the couplings
${\mathfrak a}_{n+\nu_{0}}^{(l)}$, all RG terms contributing to this
order, while the resummed $\pi^2$ terms are integral parts of the 
\emph{analytic} couplings by construction.

%%%%%%%%%%%%%%%%%%%%%%%%%%%%%%%%%%%%%%%%%%%%%%%%%%%%%%%%%%%%%%%%%%%%%%%
\subsection{Comparison of different perturbative approaches to obtain
            $\widetilde{R}_\text{S}$}
\label{subsec:various-Rs}
%%%%%%%%%%%%%%%%%%%%%%%%%%%%%%%%%%%%%%%%%%%%%%%%%%%%%%%%%%%%%%%%%%%%%%%

We list below the results obtained for the quantity
$\widetilde{R}_{\text{S}}$ using different methods.
\begin{itemize}
\item Broadhurst, Kataev, and Maxwell (BKM) \cite{BKM01} utilized 
within the so-called ``naive non-Abelianization'' (NNA) approach an 
optimized power-series expansion, based on the ``contour integration'' 
technique, to compute $\widetilde{R}_{\text{S}}$.
Their estimate, with one-loop running of $a_s$ and setting 
$a_{s}^{(l=1)}\equiv a_{s}$, reads (see Section 3.3 in 
\cite{BKM01})\footnote{The couplings $a_s$ and ${\mathfrak a}_\nu$ 
are understood to be functions of $s$, i.e., $a_s(s)$ and 
${\mathfrak a}_\nu(s)$.
Note in this context that in the original paper of Ref.\ \cite{BKM01}, 
the authors used the coefficients $d_{n}^{\text{NNA}}$, which are the 
``all-order'' coefficients of the expansion of the Adler function in 
the Euclidean region, estimated, however, through the NNA procedure.}
\begin{eqnarray}
 \widetilde{R}_{\text{S}}^{(l=1)\text{BKM}}
 &=& 3\,\hat{m}^{2}_{(l=1)}
        \left(a_{s}\right)^{\nu_{0}}
         \left[A_{0}^{\text{BKM}}(a_{s})
            + \sum_{n\geq1}^{} d_{n}\, A_{n}^{\text{BKM}}(a_{s})
         \right],
\label{eq:R-BKM2a} \\
{A}^{\text{BKM}}_n(a_{s})
 &=& \frac{4}{b_{0}\,\pi\,\delta_{n}}
   {\left[1+\left(\frac{b_{0}\,\pi\,a_{s}}{4}\right)^{2}
    \right]}^{-\delta_{n}/2}
   \left(a_{s}\right)^{n-1}{\sin}\left({\delta}_{n}
    \arctan \left(\frac{b_{0}\,\pi\,a_{s}}{4}\right)\right),~~~~~~~
\label{eq:R-BKM2b}\\
{\delta}_{n}
 &=& n+\nu_0-1\,.
\label{eq:R-BKM2c}
 \end{eqnarray}
%Eqs (4.22) (4.23) (4.24)
These new couplings ${A}^{\text{BKM}}_n(a_{s})$ and the whole result 
are closely related to our analytic approach at the one-loop level, 
as we will show shortly.
\item
Baikov, Chetyrkin, and K\"{u}hn (BChK) \cite{BCK05} have derived within 
the standard perturbative QCD at the 
${\cal O}(a_{s}^{4})$, c.f.\ Eq.\ (\ref{eq:RSnumx})), the following 
expression:
\begin{eqnarray}
  \widetilde{R}_{\text{S}}^{(l=4)\text{BChK}}
& = &
  3\bar{m}^{2}(s)^{(l=4)}
  \left[1 + \sum_{n\geq 1}^{4} r_{n}
  \left(a_{s}^{(l=4)}\right)^{n}
  \right]\, .
\label{eq:R-Che}
\end{eqnarray}
%Eq (4.25)
\item
Consider now the MFAPT equation (\ref{eq:R-MFAPT}) and recast it in the 
form
\begin{eqnarray}
 \widetilde{R}_\text{S}^{(l)\text{MFAPT}}
 \!&=&\! 3\,\hat{m}_{(l)}^2
       \left[{\mathfrak a}_{\nu_{0}}^{(l)}
        + \sum_{n\geq1}^{l}d_{n}\,
               {\mathfrak a}_{n+\nu_{0}}^{(l)}
            + \sum_{m\geq1}^{l+4} \Delta_{m}^{(l)}\,
               {\mathfrak a}_{m+\nu_{0}}^{(l)}
                   \right]
 \label{eq:R-MFAPT-2}\\
 \!&=&\! 3\,\hat{m}_{(l)}^2
       \left[{\mathfrak a}_{\nu_{0}}^{(l)}
       + \sum_{n\geq1}^{l}
                d_{n}\left({\mathfrak a}_{n+\nu_{0}}^{(l)} 
            + \sum_{m\geq1}^{4}e_m^{(l)}\,
              {\mathfrak a}_{n+m+\nu_{0}}^{(l)} \right) 
            + \sum_{m\geq1}^{4} e_{m}^{(l)}
              {\mathfrak a}_{m+\nu_{0}}^{(l)}
             \right]\!,~~~~~~~~
 \label{eq:R-MFAPT-2-m}
\end{eqnarray}
%Eq (4.26) (4.27)
where we have truncated the ``evolution series'' at $m=l+4$, adopting 
the empirical recipe discussed after Eq.\ (\ref{eq:d-tild_1}).
Pay attention that the terms ${\mathfrak a}_{m+\nu_{0}}^{(l)}$ 
contain---by means of the index $\nu_0$---\textit{all} $\gamma_0$ and 
$b_0$ terms, and also \emph{all} $\pi^2$ terms, while the contributions 
of the higher-loop RG-dependent parts are accumulated in the 
coefficients $\Delta_{m}^{(l)}$---see Eqs.\ (\ref{eq:m2-hat-run}), 
(\ref{varphi}) and also (\ref{eq:D-approx}), (\ref{eq:d-tild_1})---in 
terms of the RG-parameters $\gamma_i$ and $b_j$.
\end{itemize}
Equation (\ref{eq:R-MFAPT-2}) can be considered as a generalization 
of the BKM ``contour-improved'' expansion in 
Eqs.\ (\ref{eq:R-BKM2a})--(\ref{eq:R-BKM2c}).
To see this, one should reduce the above expression to the one-loop 
case, by setting $\Delta_{m}^{(l)} \to 0$ and
${\mathfrak a}_{\nu}^{(l)} \to {\mathfrak a}_{\nu}^{(1)}$,
with the aim to reproduce the ``contour-improved'' effective coupling
in Eq.\ (\ref{eq:R-BKM2a}).
In fact, taking into account the explicit expression
(\ref{eq:TL_ElFun}) for ${\mathfrak a}_{\nu}^{(1)}$, one can readily
find the relation
\begin{eqnarray}
{\mathfrak a}_{n+\nu_{0}}^{(1)}
 = \left(\frac{4}{b_0}\right)^{n+\nu_{0}}\,
    {\mathfrak A}_{n+\nu_{0}}^{(1)}
 = \left(a_{s}\right)^{\nu_0}  A_{n}(a_{s})
\label{eq:A-U}
\end{eqnarray}
%Eq (4.28)
recalling  (\ref{eq:R-BKM2b}) and (\ref{eq:R-BKM2c}).
This relation establishes the equivalence between the
``contour-improved'' effective coupling and the timelike analytic
coupling ${\mathfrak a}_{\nu}^{(1)}$ in the one-loop approximation of
MFAPT. 
The only difference between 
$\widetilde{R}_{\text{S}}^{(l=1)\text{BKM}}$ 
and 
$\widetilde{R}_{\text{S}}^{(l=1)\text{MFAPT}}$
stems from the summation in Eq.\ (\ref{eq:R-BKM2a}), which extends to 
all known $d_n$-coefficients, i.e., also to those up to $n=4$.

\begin{figure}[hb]
 \centerline{~\includegraphics[width=0.49\textwidth]{%%fig-fapt-che-bkm-22-eng.eps
   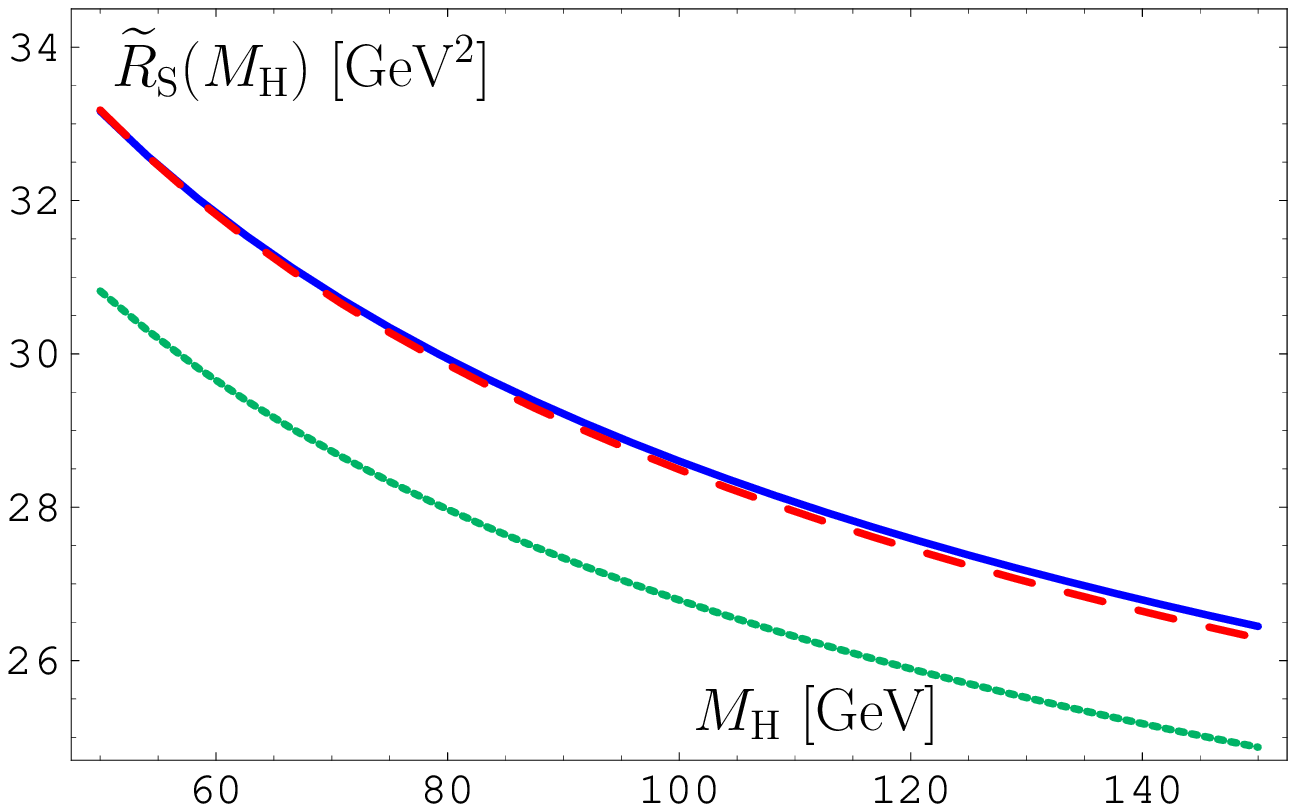}%%
             ~\includegraphics[width=0.49\textwidth]{%%fig-fapt-che-bkm-33-eng.eps
   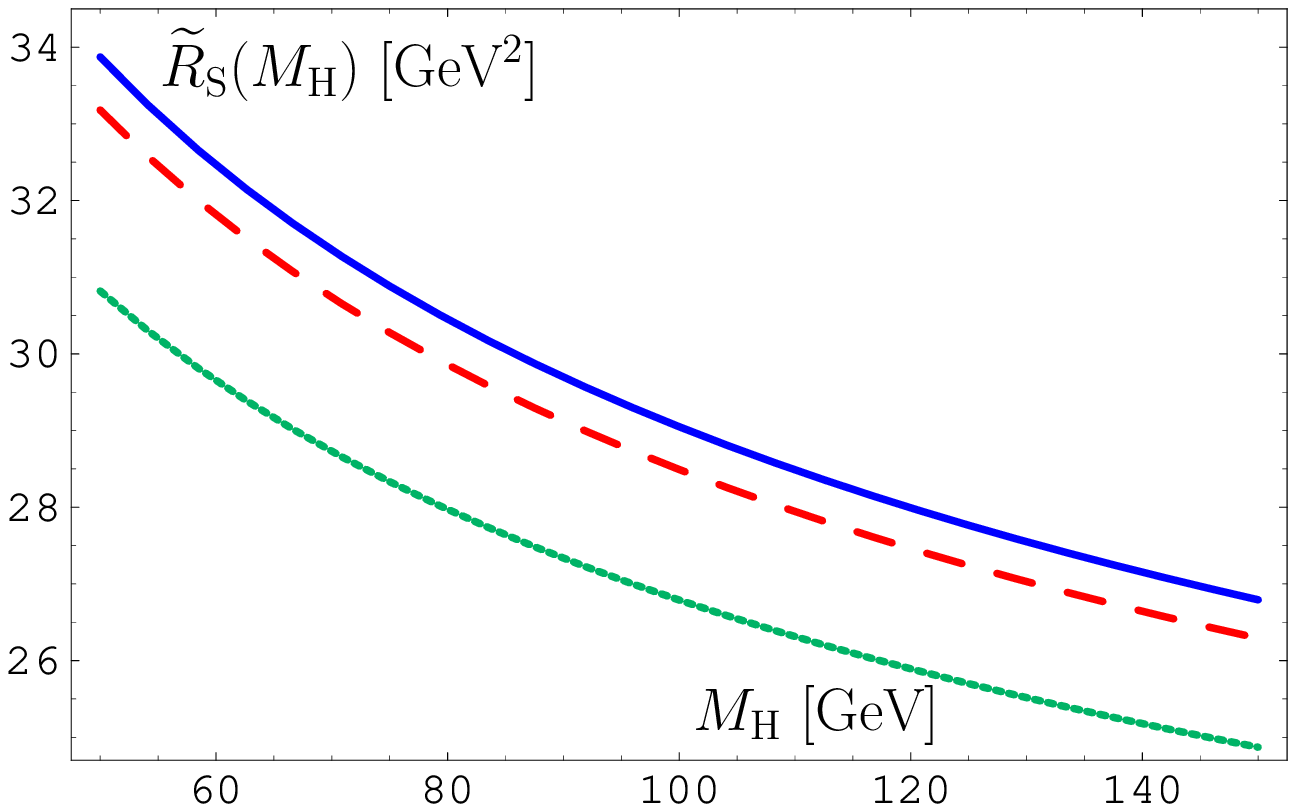}~}%
   \vspace{-0.5cm}
   \caption{Illustration of the calculation of the perturbative series
   of the quantity $\widetilde{R}_\text{S}(M^2_\text{H})$ in different
   approaches within the \protect\MS scheme: standard perturbative 
   QCD \protect\cite{ChKS97,BCK05} at the loop level $l=4$ (dashed red line, 
   $\Lambda_{N_f=5}=231$~MeV); 
   BKM estimates, by taking into account the 
   $O(\left(a_{s}\right)^{\nu_0}  A_{4}(a_{s}))$-terms, \protect\cite{BKM01} 
   (dotted green line, $\Lambda_{N_f=5}=111$~MeV); 
   and finally MFAPT from (\protect\ref{eq:R-MFAPT-2-m}) for $N_f=5$ 
   (solid blue line), displayed for $l=2$ (left panel, 
   $\Lambda_{N_f=5}=263$~MeV) and $l=3$ (right panel, 
   $\Lambda_{N_f=5}=261$~MeV).
   All $\Lambda_{N_f=5}$ scales are defined in such a way as to reproduce 
   the corresponding couplings with the value 0.120 at $s=m_Z^2$.
 \label{fig:R_S}}
\end{figure}

In Fig.\ \ref{fig:R_S}, we display the final results for the quantity
$\widetilde{R}_\text{S}^{(l)\text{MFAPT}}$ in Eq.(\ref{eq:R-MFAPT-2}) 
vs.\ the Higgs mass
$M_\text{H}$ evaluating it for the cases of two ($l=2)$ and three 
($l=3$) loops, with the goal to illustrate the effects of the 
``analytization'' procedure in comparison with the standard approach. 
The solid blue line in both panels of this figure shows the prediction 
obtained with MFAPT and a fixed number of active flavors $N_f=5$.
One appreciates that this curve lies only slightly (about $2 \%$ for 
$l=3$) above the standard result 
$\widetilde{R}_{\text{S}}^{(l=4)\text{BChK}}$,
illustrated by the dashed red line.
This is mainly due to the somewhat larger values of the perturbative 
coefficients $d_n$ within MFAPT relative to the standard ones, $r_n$.
At the same time, the dotted green line, which corresponds to 
$\widetilde{R}_{\text{S}}^{(l=1)\text{BKM}}$, Eq.\ (4.22), 
turns out to lie lower than the BChK prediction by about 7 to 8\%.
Note that if one sets $\Lambda_{N_f=5}=221$~MeV, as dictated by the 
one-loop perturbative QCD analysis of the DIS data~\cite{KPS02},
then the BKM curve, corresponding to this value, will end up to lie 
higher than the BHhK prediction by 8\%.  

%%%%%%%%%%%%%%%%%%%%%%%%%%%F I G U R E 7%%%%%%%%%%%%%%%%%%%%%%%%%%%%%%%
%%%%%%%%%%%%%%%%%%%%%%%%%%%%%%%%%%%%%%%%%%%%%%%%%%%%%%%%%%%%%%%%%%%%%%%%%%%%%%
%%%%%%                                TABLE 1                           %%%%%%
%%%%%%%%%%%%%%%%%%%%%%%%%%%%%%%%%%%%%%%%%%%%%%%%%%%%%%%%%%%%%%%%%%%%%%%%%%%%%%
\begin{table*}[t]\vspace*{-3mm}
\caption{ Comparison of the perturbative-series convergence in terms of
the associated perturbative orders, using for the Higgs boson mass the
value $M_\text{H}=120$~GeV and employing different expansions within 
the \protect\MS scheme:
(i) Standard perturbative QCD \protect{\cite{ChKS97,BCK05}},
(ii) BKM \protect{\cite{BKM01}}, and 
(iii) MFAPT at the two- and the three-loop level
of evolution with the number of flavors fixed to $N_f=5$
and $s=\left[120~\text{GeV}\right]^2$.
\label{tab:convergence}}
\begin{ruledtabular}
\begin{tabular}{lccccccc}
 Expansion approach used
     & $\widetilde{R}_\text{S}(s)$
             & $O(1)$ & $O(a_s)$& $O(a_s^2)$& $O(a_s^3)$& $O(a_s^4)$
  \\ \hline \hline
 standard QCD \protect\cite{BCK05}
     & 27.44 GeV$^2$
             & 80.2\% & 16.6\%  & 3.1\% & 0.2\% & $-0.1$\%
  \\ \hline  \hline
MFAPT
& $\widetilde{R}_\text{S}(s)$
& $O({\mathfrak a}_{\nu_{0}}^{(l)})   $ & $O({\mathfrak a}_{1+\nu_{0}}^{(l)} )$& 
  $O({\mathfrak a}_{2+\nu_{0}}^{(l)}) $ & $O({\mathfrak a}_{3+\nu_{0}}^{(l)} )$& 
  $O({\mathfrak a}_{4+\nu_{0}}^{(l)} )$
\\ \hline  \hline
  BKM \protect\cite{BKM01}\protect\footnote{Note that in this order of
 expansion, the analytic couplings in MFAPT, ${\mathfrak a}_{\nu}$, and
 those in the ``contour-improved'' approach \protect\cite{BKM01} coincide,
 though the associated coefficients $d_{n}$ and $d^{\text{NNA}}_{n}$,
 respectively, are different.} 
 (MFAPT for $l=1$) 
     & 31.89 GeV$^2$
             & 74.5\% & 17.7\%  & 5.3\% & 1.8\% & 0.7\%
  \\ \hline
  MFAPT for $l=2$, this work
     & 27.59 GeV$^2$
             & 79.5\% & 16.2\%  & 4.3\% &       &  
 \\ \hline
 MFAPT for $l=3$, this work
     & 28.07 GeV$^2$
             & 78.5\% & 16.1\%  & 4.2\% & 1.2\% &   
\end{tabular}
\end{ruledtabular}
\end{table*}
%%%%%%%%%%%%%%%%%%%%%%%%%%%%%%%%%%%%%%%%%%%%%%%%%%%%%%%%%%%%%%%%%%%%%%%%%%%%%%
%%% End-of-table %%%%%%%%%%%%%%%%%%%%%%%%%%%%%%%%%%%%%%%%%%%%%%%%%%%%%%%%%%%%%
%%%%%%%%%%%%%%%%%%%%%%%%%%%%%%%%%%%%%%%%%%%%%%%%%%%%%%%%%%%%%%%%%%%%%%%%%%%%%%

In Table \ref{tab:convergence} we show for the quantity
$\widetilde{R}_\text{S}(s)$ a comparison of the perturbative series
evaluated in different expansion approaches, discussed in the 
literature and in this work.
One may conclude that the standard perturbation-theory series and MFAPT
show a similar behavior, starting with the two-loop running.
Because the coefficients $d_n$ of MFAPT are larger than the standard 
ones, $r_n$, due to the subtraction of the $\pi^2$ terms in the latter, 
the fixed-order MFAPT result for the quantity 
$\widetilde{R}_\text{S}(s)$ appears to be slightly larger than that of 
the standard perturbative QCD approach.
However, the coupling parameters of MFAPT,
${\mathfrak a}_{\nu}^{}$, contain the resummed contribution of an 
infinite series of $\pi^2$-terms that renders them ultimately smaller 
(for the same value of $\Lambda_\text{QCD}$) than the corresponding 
powers of the standard coupling, as it can be seen from 
Table \ref{tab:evolution}.
%%%%%%%%%%%%%%%%%%%%%%%%%%%%%%%%%%%%%%%%%%%%%%%%%%%%%%%%%%%%%%%%%%%%%%%%%%%%%%
%%%%%%                                TABLE 2                          %%%%%%
%%%%%%%%%%%%%%%%%%%%%%%%%%%%%%%%%%%%%%%%%%%%%%%%%%%%%%%%%%%%%%%%%%%%%%%%%%%%%%
\begin{table*}[hb]
\caption{Comparison of the combined evolution effects on the ratio 
of MFAPT to the standard perturbation theory for the three-loop 
evolution case and fixing the number of flavors to $N_f=5$.
For the sake of completeness, we also show the ratios of the 
associated perturbative coefficients $d_n$ and $r_n$ and the 
approximate relative weights 
$\Ds\omega_n=r_n E_{n+\nu_0}^{(3)}/
 \sum_{i=0,...,3}r_i E_{i+\nu_0}^{(3)}$
of the corresponding contributions.
\label{tab:evolution}}
\begin{ruledtabular}
\begin{tabular}{ccccc}
             &~~~$n=0$~~~&~~~$n=1$~~~&~~~$n=2$~~~&~~~$n=3$~~~\\ \hline
 ~~~$\Ds\big[{\cal E}_{n+\nu_0}^{(3)}\vphantom{^\big|}/{E_{n+\nu_0}^{(3)}}\big]_{s=(120~\text{GeV})^2}$~~~
             &~~~1.00~~~&~~~0.98~~~&~~~0.95~~~&~~~0.91~~~ \\
   $d_{n}/r_{n}\vphantom{_|}$
             &~~~1.00~~~&~~~1.00~~~&~~~1.44~~~&~~~8.46~~~ \\
   relative weight $\omega_n\vphantom{_|}$
             &~~~0.801~~~&~~~0.166~~~&~~~0.031~~~&~~~0.002~~~
\end{tabular}
\end{ruledtabular}
\end{table*}
%%%%%%%%%%%%%%%%%%%%%%%%%%%%%%%%%%%%%%%%%%%%%%%%%%%%%%%%%%%%%%%%%%%%%%%%%%%%%%
%%% End-of-table %%%%%%%%%%%%%%%%%%%%%%%%%%%%%%%%%%%%%%%%%%%%%%%%%%%%%%%%%%%%%
%%%%%%%%%%%%%%%%%%%%%%%%%%%%%%%%%%%%%%%%%%%%%%%%%%%%%%%%%%%%%%%%%%%%%%%%%%%%%%
In order to demonstrate the combined effect of evolving the 
coefficients and the heavy quark mass at different orders of the 
perturbative expansion, we define the following factors for, 
respectively, the standard perturbation theory (PT) and for MFAPT:
\begin{eqnarray}
   E_{\nu}^{(l)}(s)
  &=& \left(a_s(s)\right)^{\nu}
   + \sum_{m\geq1}^{4} e_{n}^{(l)} \left(a_s(s)\right)^{m+\nu}\,;
 \label{eq:E.PT.nu}\\
  {\cal E}^{(l)}_{\nu}(s)
  &=& {\mathfrak a}_{\nu}(s)
   + \sum_{m\geq1}^{4} e_{n}^{(l)}
     {\mathfrak a}_{m+\nu}(s)\,.
 \label{eq:E.cal.nu}
\end{eqnarray}
%Eq (4.29) (4.30)
Then Eqs.\ (\ref{eq:R-Che}) and (\ref{eq:R-MFAPT-2-m}) can be rewritten 
as
\begin{eqnarray}
 \widetilde{R}_\text{S}^{(l)\text{PT}}
  = 3\,\hat{m}_{(l)}^2\,
       \left[E_{\nu_0}^{(l)}(s)
            + \sum_{n\geq1}^{l}
                r_{n}
                 E_{n+\nu_0}^{(l)}(s)
       \right]\!;
 \label{eq:R-BChK-2.E}\\
 \widetilde{R}_\text{S}^{(l)\text{MFAPT}}
  = 3\,\hat{m}_{(l)}^2\,
       \left[{\cal E}_{\nu_0}^{(l)}(s)
            + \sum_{n\geq1}^{l}
                d_{n}
                 {\cal E}_{n+\nu_0}^{(l)}(s)
       \right]\,.
 \label{eq:R-MFAPT-2.E}
\end{eqnarray}
%Eq (4.31) (4.32)
In order to understand the effect of ``analytization'', we compare the 
values of ${\cal E}_{n+\nu_0}^{(l)}(s)$ with those of 
$E_{n+\nu_0}^{(l)}(s)$, and display their ratio in 
Table \ref{tab:evolution}.
Using the entries of this Table, one can easily estimate the relative 
enhancement of 
$\widetilde{R}_\text{S}^{(3)\text{MFAPT}}$
with respect to $\widetilde{R}_\text{S}^{(3)\text{PT}}$:
\begin{eqnarray}
&& \frac{\widetilde{R}_\text{S}^{(3)\text{MFAPT}}}
      {\widetilde{R}_\text{S}^{(3)\text{PT}}}
   =
       \sum_{n\geq0}^{3}\omega_{n}\,
        \frac{d_{n}}{r_n}\,
         \frac{{\cal E}_{n+\nu_0}^{(3)}(s)}{E_{n+\nu_0}^{(3)}(s)}
   = 0.801 
   + 0.163 
   + 0.042 
   + 0.015
   = 1.021\,,               
 \label{eq:R.MFAPT.BChK}    \\
&&\mbox{where}~ \Ds\omega_n=r_n E_{n+\nu_0}^{(3)}/
  \sum_{i=0,\ldots,3}r_i E_{i+\nu_0}^{(3)}.
\end{eqnarray}
%Eq (4.33) (4.34)

From this equation in conjunction with the values of $\omega_n$ 
(presented in Table \ref{tab:evolution}), we see that the largest 
enhancement is provided for $n=3$ due to $d_n/r_n=8.46$, amounting to 
a $1.5\%$ contribution to $\widetilde{R}_\text{S}^{(3)\text{MFAPT}}$
out of $2.1\%$ in total.
Evolution, which potentially could reduce this enhancement, 
because of the inclusion 
into the analytic coupling of the resummed $\pi^2$ terms, 
notably, 
${\cal E}_{n+\nu_0}^{(3)}/{E_{n+\nu_0}^{(3)}}=0.91$,
is too small to counterbalance it.

%%%%%%%%%%%%%%%%%%%%%%%%%%%%%%%%%%%%%%%%%%%%%%%%%%%%%%%%%%%%%%%%%%%%%%%
%%%%%%%%%%%%%%%%%%%%%%%%%%%%%%%%%%%%%%%%%%%%%%%%%%%%%%%%%%%%%%%%%%%%%%%
\section{Summary and Conclusions}
 \label{sec:concl}
%%%%%%%%%%%%%%%%%%%%%%%%%%%%%%%%%%%%%%%%%%%%%%%%%%%%%%%%%%%%%%%%%%%%%%%

This report has focused on the implementation of analyticity of QCD
amplitudes both in the Euclidean and in the Minkowski space, the 
theoretical basis being provided by dispersion relations (to ensure
causality) in conjunction with the renormalization group.
The goal of the work was to elevate Analytic Perturbation
Theory---initiated by Shirkov and Solovtsov \cite{SS97}---to a
calculational paradigm for perturbative QCD applications, capable of
providing singularity-free expressions for any real power of the strong
coupling.
Following the rationale that all quantities that may contribute to the
spectral density should be included into the ``analytization'' procedure
\cite{KS01,Ste02}, we have expanded the previously developed Fractional
Analytic Perturbation Theory \cite{BMS05} from the spacelike to the
timelike region.

The core issues of the investigation can be summarized as follows,
starting with the benchmarks of the analytic couplings
${\cal A}_{\nu}^{(l)}(L)$ (spacelike)
and ${\mathfrak A}_{\nu}^{(l)}(L_s)$ (timelike) of (M)FAPT
for an arbitrary real index $\nu$:
\begin{itemize}
\item $\left\{{\cal A}_{\nu}^{(l)}(L),
      {\mathfrak A}_{\nu}^{(l)}(L_s)\right\}
      $
      is analytic in $L$ (respectively, $L_s$) and also in the index
      $\nu \in \mathbb{Z}$; for $l=1$, see Eqs.\ (\ref{eq:A-F}) and
      (\ref{eq:U-F}); for $l=2$, see Sec.\ III~C and 
      also \cite{Mag05} (for integer indices).
      This implies the existence of any index derivative
      $\Ds \frac{d^m}{d\nu^m}\,\left\{{\cal A }_{\nu}^{(l)}(L),
      {\mathfrak A}_{\nu}^{(l)}(L_s)\right\}
      $.

\item ${\cal A}_{m}^{(l)}(-\infty)=
       {\mathfrak A}_{m}^{(l)}(-\infty)=\delta_{1,m}
      $  for $ m\in \mathbb{N}$ \cite{Shi98} and
      ${\cal A}_{\nu}^{(l)}(-\infty)
      = {\mathfrak A}_{\nu}^{(l)}(-\infty)=0$
      for $\nu > 1 $.
      This implies the existence of a universal IR fixed point.
\item ${\mathfrak A}_{\nu}^{(l)}(L_s \to \infty)
       \to {\cal A}_{\nu}^{(l)}(L \to \infty)
       \to  a^{\nu}_{(l)}(L \to \infty)
      $
      have the correct UV asymptotics, dictated by the asymptotic 
      freedom of QCD.

\item The effect of the ``distorted-mirror symmetry'' with respect to
      $L$ and $L_s$, is exhibited analytically in terms of
      ${\cal A}^{(l)}_{\nu}(L)$ and ${\mathfrak A}^{(l)}_{\nu}(L_s)$ 
      for any real $\nu$.

\item The ``analytization'' of expressions, like
      ${\cal L}_{\nu,m}^{(l)}(L)
      =
      \textbf{A}_\text{E}\left[\left(a_{(l)}(L)\right)^\nu L^m\right]$,
      ${\cal L}_{\nu,m}^{(l)}(L_s)
      =\textbf{A}_\text{M}\left[\left(a_{(l)}(L)\right)^\nu L^m\right]$,
      which amount to `evolution logarithmic factors' in $l$-order QCD
      perturbation theory, typical examples being logarithms of the 
      factorization scale (appearing beyond the leading-order 
      expansion), and significantly affecting the precision of 
      perturbative calculations, becomes possible.
      Moreover, the insensitivity to the choice of the factorization 
      scale \cite{BKS05} and a diminished dependence on the adopted 
      renormalization scheme and scale setting has been achieved 
      \cite{BPSS04}.
\end{itemize}

The following topics have been given particular attention:
\begin{enumerate}
\item We have presented a brief historical review of the
``analytization'' technology from early attempts in QED up to
the most recent developments in QCD perturbation theory and
phenomenology (Introduction), highlighting the strategy for
resolving the conflict with analyticity in $Q^2$ of QCD ``observables''
in the perturbative regime.
\item In sections \ref{sec:basics} and \ref{sec:MFAPT} we have worked
out in detail the theoretical framework to deal with analytic versions
of the strong coupling and its powers for any fractional (real) power,
encompassing the spacelike and also the timelike region.
\item We have given closed-form expressions for the analytic-coupling
images at the one-loop level both in the Euclidean, as well as in the
Minkowski space.
We have also provided explicit but approximate expressions at the 
two-loop level, which show excellent agreement with the exact but 
numerical results in terms of the Lambert function, found before by 
Magradze \cite{Mag03u}---see Fig.\ \ref{fig:FAPT-MFAPT-2}.
Three-loop (approximate) results have also been included and the fast 
convergence of the non-power-series expansion has been proved.
\item The major characteristics (advantages) of this machinery, studied
in Sec.\ \ref{sec:MFAPT}, are:
(i) a reduced uncertainty level of only a few per mil in a wide range
of momentum (or energy) values, starting just above $1$~GeV,
(ii) the high quality of the non-power-series expansion, 
 evidenced by providing trustworthy analytic expressions 
 for the strong coupling and its powers 
 at the one, two- and three-loop level that 
 are singularity free in the spacelike region 
 and resum the $\pi^2$ terms (induced by analytic continuation) 
 to all orders in the timelike domain.
\item The relevance of the developed framework 
 for practical purposes in the Minkowski space 
 has been effected in Sec.\ \ref{sec:higgs} 
 by applying it 
 to the decay of a scalar Higgs boson to a $b\bar{b}$ pair, 
 this task serving as a Proof-of-the-Concept-calculation.
Specifically, we estimated the quantity $R_\text{S}$ at the four-loop 
level of the perturbative expansion (i.e., up to the coefficient 
$d_3$), having recourse to the coefficients and anomalous dimensions 
calculated by Chetyrkin and collaborators in \cite{Che96,ChKS97}.
The main advantage of MFAPT here is that the coupling parameters
${\mathfrak A}_{\nu}$ inherently include the resummed contribution of 
an infinite series of those $\pi^2$-terms originating from the 
analytic continuation from the Euclidean to the Minkowski space.
Technically, we could have also included the next correction, 
associated with the coefficient $d_4$, but this would be not worth the 
effort, given that the expected contribution would be about $0.5\%$.
\end{enumerate}

In conclusion, we think that the analytic approach to QCD perturbation
theory in the form advocated here has indeed been rather successful
both in the spacelike and in the timelike region for the particular
processes we have discussed in this work and elsewhere
\cite{BMS05,BKS05,BPSS04}.
The developed full-fledged machinery may 
improve our understanding of key issues of QCD reactions 
from the poorly understood low-momentum (spacelike) domain, 
where the Landau singularities 
are a serious obstacle in standard perturbative QCD 
(based on power-series expansions) up to the high-energy 
(timelike) region of several GeV.
In conjunction with high-loop calculations in the latter case, 
the $\pi^2$ terms, entailed by analytic continuation, can
significantly change the result in higher orders.
A greater challenge, however, is to include into the approach 
gluon resummation and power corrections 
analytically, 
beyond the attempts in \cite{KS01} and \cite{BMS05}, 
with the aim to unravel the analytic structure 
of the involved amplitudes in the whole momentum (energy) range.
Work in this direction is currently in progress. 

%%%%%%%%%%%%%%%%%%%%%%%%%%%%%%%%%%%%%%%%%%%%%%%%%%%%%%%%%%%%%%%%%%%%%%%
\acknowledgments
%%%%%%%%%%%%%%%%%%%%%%%%%%%%%%%%%%%%%%%%%%%%%%%%%%%%%%%%%%%%%%%%%%%%%%%
We would like to thank Yakov I.\ Azimov and Dmitry V.\ Shirkov 
for stimulating discussions and useful remarks.
We wish to thank A.\ L.\ Kataev for useful remarks pertaining 
to Fig.\ 7. 
N.G.S is grateful to BLTPh@JINR for support, where much of this work
was carried out. 
Two of us (A.P.B.\ and S.V.M.) are indebted to Prof.\ Klaus Goeke for 
the warm hospitality at Bochum University, where this work was 
completed.
This work was supported in part by the Deutsche Forschungsgemeinschaft
(Project DFG 436 RUS 113/881/0-1),
the Heisenberg--Landau Programme, grant 2006, 
the Russian Foundation for Fundamental Research, 
grants No.\ 05-01-00992 and No.\ 06-02-16215,
and the BRFBR--JINR Cooperation Programme, 
contract No.\ F06D-002.

\begin{appendix}
\appendix
%%%%%%%%%%%%%%%%%%%%%%%%%%%%%%%%%%%%%%%%%%%%%%%%%%%%%%%%%%%%%%%%%%%%%%%
%%%%%%%%%%%%%%%%%%%%%%%%%%%%%%%%%%%%%%%%%%%%%%%%%%%%%%%%%%%%%%%%%%%%%%%
\section{Two-loop renormalization-group solutions for the coupling }
 \label{RG-solution}
%%%%%%%%%%%%%%%%%%%%%%%%%%%%%%%%%%%%%%%%%%%%%%%%%%%%%%%%%%%%%%%%%%%%%%%

\textbf{1.} The expansion of the $\beta$-function is given by the RHS 
of the equation 
\begin{eqnarray}
 \frac{d}{dL}\left(\frac{\alpha_{s}}{4 \pi}\right)
=
  \beta\left( \frac{\alpha_{s}}{4 \pi}\right)
= - b_0\left(\frac{\alpha_{s}}{4 \pi}\right)^2
  - b_1\left(\frac{\alpha_{s}}{4 \pi}\right)^3
  - b_2\left(\frac{\alpha_{s}}{4 \pi}\right)^4\,- \ldots \, ,
\label{eq:betaf}
\end{eqnarray}
%Eq (A1)
where $L=\ln(\mu^2/\Lambda^2)$ and
\begin{eqnarray}
    b_0 &=& \frac{11}{3}\,C_\text{A} - \frac{4}{3}\,T_\text{R} N_f
    \,;\qquad 
    b_1 = \frac{34}{3}\,C_{\text{A}}^{2}
        - \left(4C_\text{F}
        + \frac{20}{3}\,C_\text{A}\right)T_\text{R} N_f ;\nonumber \\
    b_2 &=&   \frac{2857}{54} C_\text{A}^3
 +2 C_\text{F}^2 T_\text{R} N_f - \frac{205}{9} C_\text{F} C_\text{A} T_\text{R} N_f 
 - \frac{1415}{27} C_\text{A}^2 T_\text{R} N_f 
 + \frac{44}{9} C_\text{F} (T_\text{R} N_f)^2 \nonumber \\
 && 
 + \frac{158}{27} C_\text{A} (T_\text{R} N_f)^2\,,  
 \label{eq:beta0&1&2}
\end{eqnarray}
%Eq (A2)
with $C_\text{F}=\left(N_\text{c}^{2}-1\right)/2N_\text{c}=4/3$,
$C_\text{A}=N_\text{c}=3$, $T_\text{R}=1/2$, and $N_f$ denoting
the number of active flavors.
The corresponding three-loop RG equation for the coupling
$a=b_0\,\alpha/(4\pi)$ is
\begin{eqnarray}
 \frac{d a_{(3)}}{dL}
  = - a_{(3)}^2\left[1 + c_1\,a_{(3)}+ c_2\,a_{(3)}^2\right]
 \text{~~with~}c_1
 \equiv
 \frac{b_1}{b_0^2}, ~c_2\equiv\frac{b_2}{b_0^3} \, .
\label{eq:beta.new}
\end{eqnarray}
%Eq (A3)
The solution of this RG equation at the two-loop level ($c_2=0$) 
assumes the form 
\begin{eqnarray}
 \label{eq:App-RGExact}
 \frac{1}{a_{(2)}} +
 c_1
     \ln\left[\frac{a_{(2)}}{1+c_1 a_{(2)}}
     \right] = L\,.
\end{eqnarray}
%Eq (A4)
The exact solution of Eq.\ (\ref{eq:App-RGExact}) can be expressed in
terms of the Lambert function $W(z)$, \cite{CGHJK96} (see also
\cite{Mag99,Mag00}) defined by
\begin{eqnarray}
z=W(z)\, e^{W(z)}\,.
\end{eqnarray}
%Eq (A5)
This solution has the form
\begin{eqnarray}
 \label{eq:App-Exactsolution}
 a_{(2)}(L) =
 -\frac{1}{c_1} \frac{1}{1+W_{-1}(z(L))}\, ,
 \end{eqnarray}
%Eq (A6)
where
$z(L) =
\left(1/c_1\right) \exp\left(-1+i\pi-L/c_1\right)
$
and the branches of the  multivalued function $W$ are denoted by
$W_{k}$, $k=0,\pm 1,\ldots $.
A review of the properties of this special function can be found in
\cite{CGHJK96}; see also \cite{Mag99} for a special emphasis on the
problem considered here.

\textbf{2.}~The expansion of the solution of Eq.\ (\ref{eq:beta.new}), 
i.e., $a_{(3)}(L)$, in terms of $a_{}=1/L$, while retaining terms of 
order $a_{}^{4}$, yields
\begin{eqnarray}
 \label{eq:Delta.PT}
  a_{(3)}
   &=& a_{}
     + a_{}^2\, c_1\,\ln a_{(1)}
     + a_{}^3\,
        \left[c_1^2\, \left(\ln^2 a_{}
            + \ln a_{}
             -1 \right) +c_2\right]  \nonumber \\
&& ~~+a_{}^4\,
        \left[ c_1^3\, \left(\ln^3 a_{}
            + \frac5{2}\ln^2 a_{}
            -2~\ln a_{}
            -\frac1{2} \right) +3 c_2 c_1\ln a_{}\right]\nonumber \\
 &&            ~~+ {\cal O}\left(a_{}^{5}\ln^{4} a_{}\right)\,.
\end{eqnarray}
%Eq (A7)
For any power $\nu$ of the coupling we have the final three-loop 
expression
\begin{eqnarray}
 \left(a_{(3)}\right)^\nu
 &=& a_{}^{\nu}
    + c_1\nu\,a_{}^{\nu+1}\ln a_{}
    + \left[c_1^2\left(\frac{\nu +1}{2\,!}\,\ln^2a_{}
                      + \ln a_{} - 1
                 \right)
          + c_2
      \right]\nu\,a_{}^{\nu +2} \nonumber \\
 &+&  c_1^3\nu\,a_{}^{\nu+3}
       \left[\frac{(\nu +1)(\nu +2)}{3 !}\,\ln^3a_{}
                     + \frac{2\,\nu+3}{2}\,\ln^2a_{}
                     - (\nu+1)\,\ln a_{}
                     - \frac1{2}
       \right] \nonumber \\
 &+&  c_1c_2\nu(\nu+2)\,a_{}^{\nu+3}\ln a_{} \nonumber \\
 &+&  c_1^4\nu\,a_{}^{\nu+4}
        \left[\frac{(\nu +1)(\nu +2)(\nu +3)}{4 !}\,\ln^4 a_{}
            + \frac{3\,\nu^2+12\,\nu+11}{6}\,\ln^3 a_{}
        \right. \nonumber \\
 && ~~~~~~~~~~~
        \left.
            - \frac{\nu^2+2\,\nu}{2}\,\ln^2 a_{}
            - \frac{3\,\nu+5}{2}\,\ln a_{}
            + \frac{3\,\nu+4}{6}
        \right] \nonumber \\
 &+&  c_2\nu\,a_{}^{\nu+4}
      \left[(\nu+4)c_2\,\frac{3\nu+7}{6} 
            + c_1^2\,(\nu+2)\,
               \left(\frac{\nu+3}{2}\,\ln^2 a_{}
                    + \ln a_{} 
                    - 1 
               \right)
      \right]
 \nonumber \\
 &+& {\cal O}\left(a_{}^{\nu+5}\ln^{5} a_{}\right)\,.
  \label{eq:A8}
\end{eqnarray}
%Eq (A8)
Note that the two-loop expansion can be immediately obtained from 
Eq.\ (\ref{eq:A8}) by setting $c_2=0$.

\textbf{3.}~For completeness, we also present here the expansion of 
the product $\left[a_{(2)}\right]^{\nu} L$ in terms of $a$  that can 
be obtained (in the two-loop approximation) from 
Eq.\ (\ref{eq:App-RGExact}) :
\begin{eqnarray}
 \label{eq:a_MS_RG}
  \left(a_{(2)}\right)^\nu
   L
   = \left(a_{(2)}\right)^{\nu-1}
   + \left(a_{(2)}\right)^\nu
      c_1\,\ln\left(\frac{a_{(2)}}
                         {1+c_1a_{(2)}}
              \right)\,.
\end{eqnarray}
%Eq (A9)
Expanding the logarithmic term $\ln\left(1+c_1a_{(2)}\right)$, while
retaining terms of order
$a_{(2)}^{\nu-1}$,
$a_{(2)}^{\nu}\ln(a_{(2)})$, $a_{(2)}^{\nu+1}$,
$a_{(2)}^{\nu+2}$,
we get
\begin{eqnarray} \label{eq:a2Ltoa1}
 \left(a_{(2)}\right)^\nu L
  &=& \left(a_{(2)}\right)^{\nu-1}
   + c_1\,\left(a_{(2)}\right)^\nu
     \ln a_{(2)}
   - c_1^2\,a_{(2)}^{\nu+1}
   + \frac{c_1^3}{2}\,a_{(2)}^{\nu+2}
   + {\cal O}\left(a_{(2)}^{\nu+3}\right)
\end{eqnarray}
%Eq (A10)
and, finally, expanding the coupling $a_{(2)}$ in terms of $a=a_{(1)}$,
\begin{eqnarray}
 \label{eq:A10}
 \left(a_{(2)}\right)^\nu L
  &=& a_{}^{\nu-1}
    + c_1\,\nu\,a_{}^\nu\,\ln a_{}
    + c_1^2\,\nu~a_{}^{\nu +1} \left[\frac{\nu +1}{2}\,\ln^2(a_{})
    + \ln(a_{}) -1\right] \nonumber \\
 && ~~~~~\,+\ {\cal O}\left(a^{\nu+3}\,\ln^3 a\right)
\end{eqnarray}
%Eq (A11)
in accordance with Eq.\ (\ref{eq:A8}).
An analogous expression can be constructed for the three-loop case:
\begin{eqnarray}
 \left(a_{(3)}\right)^\nu L
 &=& a_{}^{\nu-1}
    + c_1\nu\,a_{}^{\nu}\ln a_{}
    + \left[c_1^2\left(\frac{\nu +1}{2\,!}\,\ln^2a_{}
                      + \ln a_{} - 1
                 \right)
          + c_2
      \right]\nu\,a_{}^{\nu +1} \nonumber \\
 &+&  c_1^3\nu\,a_{}^{\nu+2}
       \left[\frac{(\nu +1)(\nu +2)}{3 !}\,\ln^3a_{}
                     + \frac{2\,\nu+3}{2}\,\ln^2a_{}
                     - (\nu+1)\,\ln a_{}
                     - \frac1{2}
       \right] \nonumber \\
 &+&  c_1c_2\nu(\nu+2)\,a_{}^{\nu+2}\ln a_{} \nonumber \\
 &+&  c_1^4\nu\,a_{}^{\nu+3}
        \left[\frac{(\nu +1)(\nu +2)(\nu +3)}{4 !}\,\ln^4 a_{}
            + \frac{3\,\nu^2+12\,\nu+11}{6}\,\ln^3 a_{}
        \right. \nonumber \\
 && ~~~~~~~~~~~
        \left.
            - \frac{\nu^2+2\,\nu}{2}\,\ln^2 a_{}
            - \frac{3\,\nu+5}{2}\,\ln a_{}
            + \frac{3\,\nu+4}{6}
        \right] \nonumber \\
 &+&  c_2\nu\,a_{}^{\nu+3}
      \left[(\nu+4)c_2\,\frac{3\nu+7}{6} 
            + c_1^2\,(\nu+2)\,
               \left(\frac{\nu+3}{2}\,\ln^2 a_{}
                    + \ln a_{} 
                    - 1 
               \right)
      \right]
 \nonumber \\
 &+& {\cal O}\left(a_{}^{\nu+4}\ln^{5} a_{}\right)\,.
  \label{eq:A8.3L}
\end{eqnarray}
%Eq (A12)

\textbf{4.} We consider here the solution of the renormalization-group 
equation in the so-called Pade-modification of the three-loop 
approximation in QCD \cite{KM01,KM03}, where the $\beta$-function of 
Eq.\ (\ref{eq:betaf}) is given by
\begin{eqnarray}
 \beta_{(3\text{-P})}\left(\frac{\alpha_{s}}{4\pi}\right)
  = - b_0\left(\frac{\alpha_{s}}{4 \pi}\right)^2
      \left[1
         +  \frac{b_1\left(\alpha_{s}/(4\pi)\right)}
                 {b_0\left(1-b_2\alpha_{s}/(4b_1\pi)\right)}
     \right]\,.
\label{eq:betaf-3L}
\end{eqnarray}
%Eq (A13)
The corresponding three-loop RG equation (\ref{eq:beta.new}) is 
modified to read
\begin{eqnarray}
 \frac{d a_{(3\text{-P})}}{dL}
  = - a_{(3\text{-P})}^2
      \left[1 
      + \frac{c_1\,a_{(3\text{-P})}}
             {1-(c_2/c_1)\,a_{(3\text{-P})}}
      \right]\,.
\label{eq:beta.new.3L}
\end{eqnarray}
%Eq (A14)
The solution of this RG equation assumes the form 
\begin{eqnarray}
 \label{eq:App-RGExact.3L}
  \frac{1}{a_{(3\text{-P})}}
    + c_1 \ln\left[\frac{a_{(3\text{-P})}}{1
    + \left(c_1-c_2/c_1\right)a_{(3\text{-P})}}\right]
  = L\,,
\end{eqnarray}
%Eq (A15)
which is very similar to Eq.\ (\ref{eq:App-RGExact}).
For this reason, the exact solution of Eq.\ (\ref{eq:App-RGExact.3L}) 
can also be given in terms of the Lambert function $W(z)$.
The solution is
\begin{eqnarray}
 \label{eq:App-Exactsolution.3L}
 a_{(3\text{-P})}(L) =
 -\frac{1}{c_1}\,
   \frac{1}{1-c_2/c_1^2+W_{-1}\big(z^{(3\text{-P})}(L)\big)}\, ,
 \end{eqnarray}
%Eq (A16)
where
$z^{(3\text{-P})}(L)
=
\left(1/c_1\right) \exp\left[-1+i\pi+c_2/c_1^2-L/c_1\right]$.
The relative accuracy of this solution, 
as compared with the original Eq.\ (\ref{eq:beta.new}) solved 
numerically, is better than 1\% for $L\geq 7$ (and better than 
$0.5\%$ for $L\geq9$).

%%%%%%%%%%%%%%%%%%%%%%%%%%%%%%%%%%%%%%%%%%%%%%%%%%%%%%%%%%%%%%%%%%%%%%%
%%%%%%%%%%%%%%%%%%%%%%%%%%%%%%%%%%%%%%%%%%%%%%%%%%%%%%%%%%%%%%%%%%%%%%%
\section{Spectral density at higher-loop level}
 \label{app:spectr}
%%%%%%%%%%%%%%%%%%%%%%%%%%%%%%%%%%%%%%%%%%%%%%%%%%%%%%%%%%%%%%%%%%%%%%%

\textbf{1.} We consider here the spectral density $\rho_{\nu}(\sigma)$
beyond the one-loop approximation.
At the $l$-loop level, $\rho_{\nu}^{(l)}(\sigma)$ can always be
presented in the same form as for the leading order, given in
Eq.\ (\ref{eq:rho-A-dis}), i.e.,
\begin{eqnarray}
  \rho_{\nu}^{(l)}(\sigma)
=
  \frac{1}{\pi}\,
  \textbf{Im}\,\big[a^{\nu}_{(l)}(-\sigma)\big]
= \frac{1}{\pi}\,\frac{\sin[\nu~
  \varphi_{(l)}(\sigma)]}{\left(R_{(l)}(\sigma)\right)^{\nu}},
\label{eq:spec-dens-nu-C2}
\end{eqnarray}
%Eq (B1)
where now the phase $\varphi_{(l)}$ and the radial part $R_{(l)}$ have
a multi-loop content.
At the two-loop level, one has to consider the imaginary part of the
Lambert function $W_{-1}$ (cf. Eq.\ (\ref{eq:App-Exactsolution})); see
later discussion.
We consider first the known second-order iterative solution 
of Eq.\ (\ref{eq:App-RGExact}) 
that provides us, with sufficient accuracy, 
the following result:
\begin{eqnarray}
  \frac{1}{a_{(2)}(L)} \to  \frac{1}{a_{(2)}^\text{it-2} (L)}
=
  L + c_1\ln\left[L + c_1 + c_1\ln\left(L + c_1 \right)\right]\, .
\end{eqnarray}
%Eq (B2)
For the approximate solution $a_{(2)}^\text{it-2}$, we have
\begin{eqnarray}
 R_{(2)}^\text{it-2}(\sigma)
  &=& \sqrt{\left[ L_\sigma
      + c_1\ln R(\sigma)
            \right]^2
      + \left[\pi
      + c_1\arccos\left(\frac{L_\sigma+c_1+c_1\ln r(\sigma)}{R(\sigma)}
                            \right)
            \right]^2
          }\,,
\label{eq:R2}
  \\
%Eq (B3)
 \varphi_{(2)}^\text{it-2}(\sigma)
  &=&\arccos\left(\frac{L_\sigma + c_1\ln R(\sigma)}
                       {R_{(2)}^\text{it-2}(\sigma)}
            \right)\,,
\label{eq:varphi2}
 \end{eqnarray}
%Eq (B4)
where
\begin{eqnarray}
 R(\sigma)
 &=& \sqrt{\left[ L_\sigma
                + c_1
                + c_1\ln r(\sigma)
           \right]^2
         + \left[\pi+c_1 \phi(\sigma)
           \right]^2
         }\, ,\\
 r(\sigma)
 &=& \sqrt{\left[L_\sigma+c_1\right]^2
          + \pi^2
          };~~~
 \phi(\sigma)\
  =\ \arccos\left(\frac{L_\sigma+c_1}{r(\sigma)}\right)
\label{eq:rho2}
\end{eqnarray}
%Eq (B5) (B6)
with $L_\sigma =\ln\left(\sigma/\Lambda^2\right)$.
The spectral density 
\begin{equation}
  \rho_{\nu=1}^{(l=2)\text{it-2}}(\sigma)
  = \frac{1}{\pi}\,\frac{\sin\left(\varphi_{(2)}^{\text{it-2}}(\sigma)\right)}
                        {R_{(2)}^{\text{it-2}}(\sigma)},
\label{eq:rho2-app}
\end{equation}
%Eq (B7) 
with the phase $\varphi_{(2)}^\text{it-2}(\sigma)$ and the radial part
$R_{(2)}^\text{it-2}(\sigma)$ from Eqs.\
(\ref{eq:R2})--(\ref{eq:rho2}), appears to be very close to the
\emph{exact}, but  numerical result for
$\rho^\text{(2)}_{1}(\sigma)$,
based on the Lambert function $W_{-1}$---see, e.g., \cite{SS99}.

%%%%%%%%%%%%%%%%%%%%%%%%%%%%%%%%%%%%%%%%%%%%%%%%%%%%%%%%%%%%%%%%%%%%%%%
%%%%%%%%%%%%%%%%%%%%%%%%%%%%%% F I G U R E  8 %%%%%%%%%%%%%%%%%%%%%%%%%
%%%%%%%%%%%%%%%%%%%%%%%%%%%%%%%%%%%%%%%%%%%%%%%%%%%%%%%%%%%%%%%%%%%%%%%
\begin{figure}[t]
 \centerline{\includegraphics[width=0.42\textwidth]{%fig-sp-den_2it.eps
   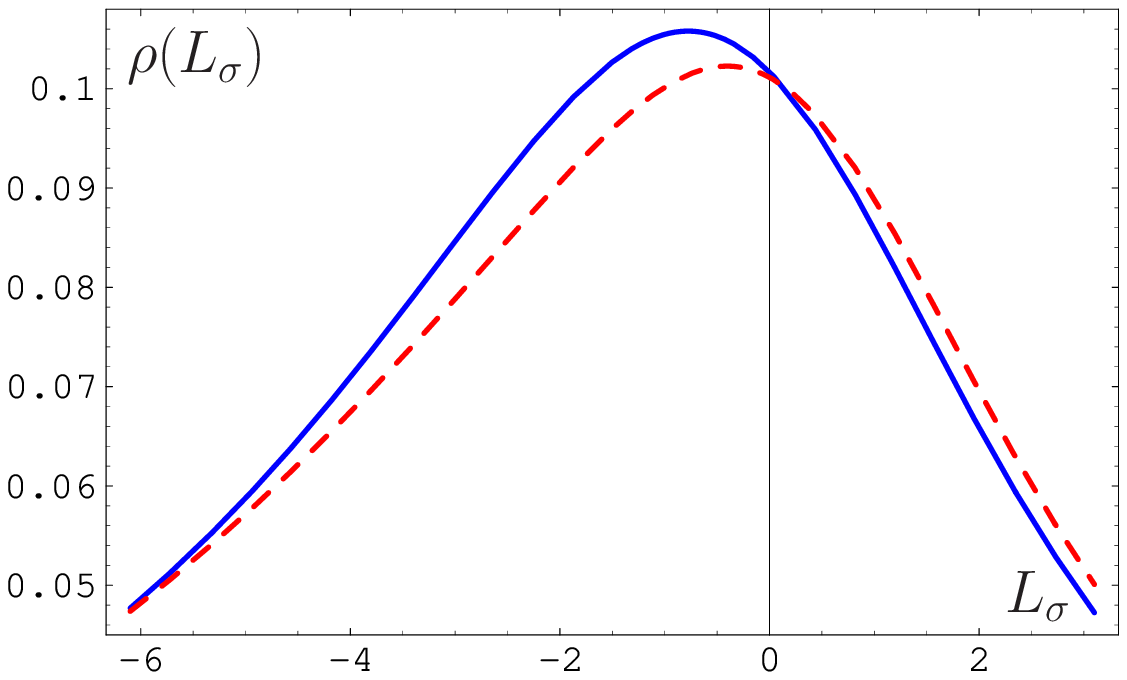}~~~%
   \includegraphics[width=0.42\textwidth]{%fig-sp-den_3it.eps
    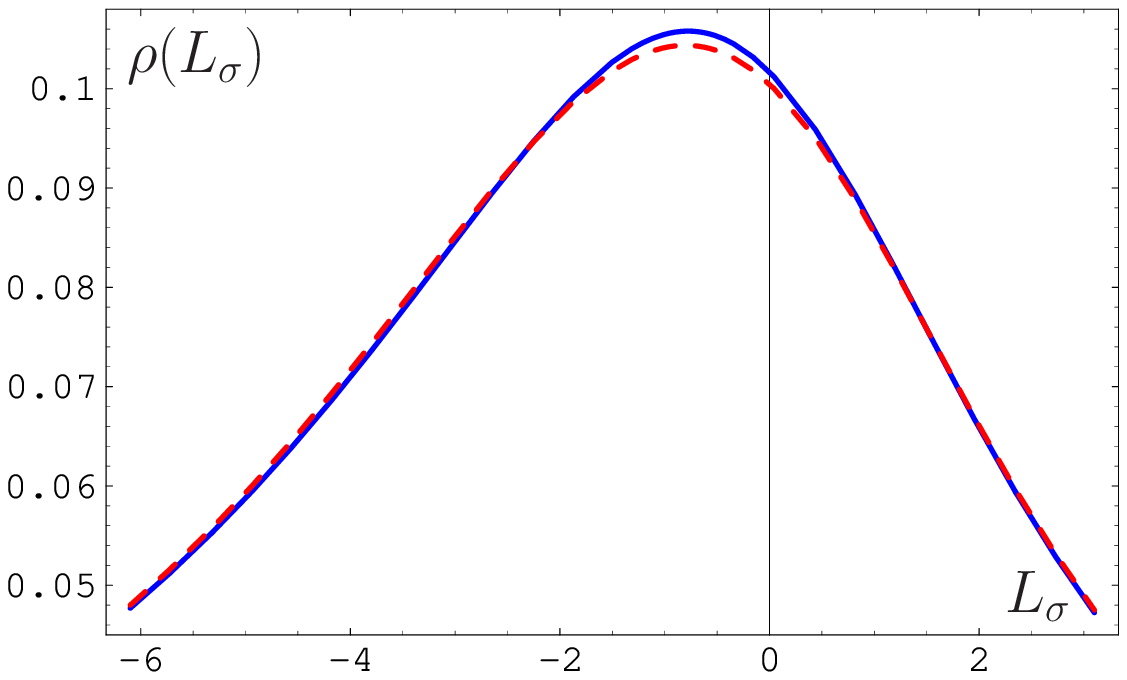}}%
   \caption{\label{fig:exact-2it}\footnotesize
   The dashed red line corresponds on the left panel to the first
   running-coupling iteration, entering the spectral density,
   $\rho^\text{(2)-it-1}_{1}(L_\sigma)$, see Eqs.\ (B12)--(B14)
   in~\cite{BMS05}, and, on the right panel, to the second
   running-coupling iteration,
   $\rho^\text{(2)-it-2}_{\nu=1}(L_{\sigma})$,
   while the solid blue line on both panels corresponds to the
   exact spectral density
   $\rho^{(2)}_1(L_\sigma)$, see Eq.\ (\ref{eq:Spec.Den.Lamb}).
   For the sake of a better comparison, the displayed region of $L_\sigma$
   is chosen in the vicinity of the maximum difference of these two
   curves.}
\end{figure}
%%%%%%%%%%%%%%%%%%%%%%%%%%%%%%%%%%%%%%%%%%%%%%%%%%%%%%%%%%%%%%%%%%%%%%%%%%%

Specifically, one can use the symbolic program
Mathematica\footnote{In versions 3, 4, and 5 of Mathematica the function
$W_{k}(z)$ is denoted by the name ProductLog$[k,z]$.},
or Maple~7, which both recognize the Lambert function, to carry out these
integrations using the spectral density \cite{Mag05}
\begin{eqnarray}
 \label{eq:Spec.Den.Lamb}
  \rho^{(2)}_m(\sigma)
   = \frac{1}{\pi}\,
      \textbf{Im}
       \left[\frac{-1}{c_1\,\left(1+W_{-1}(z_\sigma)\right)}\right]^m,
\end{eqnarray}
%Eq (B8)
where $z_\sigma=c_1^{-1}\exp\left(\Ds -L_\sigma/c_1-1+i\pi \right)$.
We present the approximate expression
$\rho^\text{(2)it-2}_{1}(L_\sigma)$ (dashed red line)
in comparison with the exact expression
$\rho^{(2)}_1(L_\sigma)$ (solid blue line)
in Fig.\ \ref{fig:exact-2it}.

\textbf{2.} It is worth noting here that explicit expressions for the 
analytic images of the integer powers of the couplings in the Minkowski 
region
\begin{eqnarray}
 \label{eq:Mink.U.Int.2Loop}
  {\mathfrak A}_{n}^{(2)}\left(L_s\right)
  &=& \int_{L_s}^{\infty}\!\!
       \rho^{(2)}_{n}(\sigma)
        dL_\sigma
\end{eqnarray}
%Eq (B9)
have been obtained before by Magradze~\cite{Mag05}, using the properties 
of the Lambert function; notably,
\begin{eqnarray}
 \label{eq:Mink.U1.2Loop}
  {\mathfrak A}_{1}^{(2)}\left(L_s\right)
  &=& 1
   - \frac{1}{\pi}\,
      \textbf{Im}\,\left[\ln{W_{1}(z_{s})}\right]\,;
  \\
 \label{eq:Mink.U2.2Loop}
   {\mathfrak A}_{2}^{(2)}\left[L_s\right]
  &=& \frac{1}{c_1\pi}\,
   \textbf{Im}\,\left[\ln\left(\frac{W_{1}\left(z_{s}\right)}
                           {1+W_{1}\left(z_{s}\right)}\right)
                \right]\,;\\
 \label{eq:Mink.Un.2Loop}
  {\mathfrak A}_{n+2}^{(2)}\left(L_s\right)
  &=& -\,\frac{1}{c_{1}}\,
        \left[{\mathfrak A}_{n+1}^{(2)}\left(L_s\right)
      + {1\over n}\,
      \frac{\mathrm{d}\phantom{L_s\!\!\!}}{\mathrm{d}L_s\!}\
           {\mathfrak A}_{n}^{(2)}\left(L_s\right)
        \right]\, \text{~for~} n\geq1\,.
\end{eqnarray}
%Eqs (B10) (B11) (B12)

\textbf{3.} For the Pade-modification of the three-loop approximation 
the corresponding spectral densities are defined through 
Eq.\ (\ref{eq:App-Exactsolution.3L})
\begin{eqnarray}
 \label{eq:Spec.Den.Lamb.3L}
  \rho^{(3\text{-P})}_m(\sigma)
   = \frac{1}{\pi}\,
     \textbf{Im}
     \left[\frac{-1}{c_1\,
     \left(1-c_2/c_1^2+W_{-1}\big(z^{(3\text{-P})}(L_\sigma)\big)
     \right)}
     \right]^m\,.
\end{eqnarray}
%Eq (B13)
The following explicit expression for the analytic image of the 
coupling in the Minkowski region 
\begin{eqnarray}
 \label{eq:U_1.3L}
  {\mathfrak A}_{1}^{(3\text{-P})}\left(L_s\right)
   = \frac{1}{\pi}
     \left\{\pi 
   - \frac{c_1^2}{c_1^2-c_2}\,
     \textbf{Im}\left[\ln{W_{1}\left(z_s\right)
     \vphantom{\frac{c_2}{c_1^2}}}
                \right]
   + \frac{c_2}{c_1^2-c_2}\,
     \textbf{Im}\left[\ln\left(1-\frac{c_2}{c_1^2}+W_1\left(z_s\right)
                         \right)
                \right]
      \right\}~~~~
\end{eqnarray}
%Eq (B14)
with $z_s=z^{(3\text{-P})}(L_s)$
has been obtained before by Magradze~\cite{Mag00}.
The relative accuracy of this solution, as compared with the numerical 
integration of the original, non-Pade-modified, spectral density 
$\rho^{(3)}_1(\sigma)$ is better than $0.25\%$ for $L\geq 2$.

\textbf{4.} A useful formula to relate arbitrary powers of logarithms
by means of the dispersion relation (\ref{eq:D-operation}) has been
presented in \cite{BKM01}; viz.,
\begin{equation}
 Q^2\int_0^\infty{ds\over(s+Q^2)^2}\left({\mu^2\over s}\right)^\delta
 = {\pi\delta\over\sin(\pi\delta)}
    \left({\mu^2\over Q^2}\right)^\delta\,.
 \label{pi2}
\end{equation}
%Eq (B15)

%%%%%%%%%%%%%%%%%%%%%%%%%%%%%%%%%%%%%%%%%%%%%%%%%%%%%%%%%%%%%%%%%%%%%%%
%%%%%%%%%%%%%%%%%%%%%%%%%%%%%%%%%%%%%%%%%%%%%%%%%%%%%%%%%%%%%%%%%%%%%%%
\section{``Analytization'' of powers of the coupling multiplied by
           logarithms}
 \label{app:AnB}
%%%%%%%%%%%%%%%%%%%%%%%%%%%%%%%%%%%%%%%%%%%%%%%%%%%%%%%%%%%%%%%%%%%%%%%

\textbf{1.}
Here we derive the ``analytization'' of  powers, 
$\left(a_{(2)}\right)^\nu$, and more complicated expressions that 
contain powers of the running coupling multiplied by logarithms, 
making use of the property
\begin{eqnarray} \label{eq:log(a)}
 \left[ a^\nu\ln(a)\right]_\text{an}
  &\equiv&
    \left[ \frac{d}{d \nu} a^\nu \right]_\text{an}
  \,\stackrel{def}{=}\,
    \frac{d}{d \nu}\,{\cal A}_{\nu}
  \,=\,{\cal D}\,{\cal A}_{\nu}\,,
\end{eqnarray}
%Eq (C1)
supplemented by Eqs.\ (\ref{eq:Delta.PT}) and (\ref{eq:A8}).
In this way we obtain
($c_1\equiv b_1/b_0^2, ~c_2\equiv b_2/b_0^3$)
\begin{subequations}
 \begin{eqnarray}
  \label{eq:image.u.2.nu}
  {{\cal A }_{\nu}^{(3)}\choose{\mathfrak A}_{\nu}^{(3)}}
   = {{\cal A }_{\nu}^{(1)}\choose{\mathfrak A}_{\nu}^{(1)}}
     + c_1\,\nu\,{\cal D}\,
       {{\cal A }_{\nu+1}^{(1)}\choose{\mathfrak A}_{\nu+1}^{(1)}}
     + \nu\left[c_1^2\,
                \left(\frac{\nu+1}{2!}\,{\cal D}^2
                    + {\cal D} - 1
                \right)
              + c_2  
           \right]
       {{\cal A }_{\nu+2}^{(1)}\choose{\mathfrak A}_{\nu+2}^{(1)}}
  ~~~~~~~~\\
  \label{eq:image.u.3.nu}
  +\,c_1 \nu
     \left\{c_1^2\left[\frac{(\nu+1)(\nu+2)}{3!}\,{\cal D}^3
                        + \frac{2\,\nu+3}{2}\,{\cal D}^2
                        + (\nu+1)\,{\cal D}
                        - \frac12
                  \right]
           + c_2(\nu+2)\,{\cal D}
     \right\}
         {{\cal A }_{\nu+3}^{(1)}\choose{\mathfrak A}_{\nu+3}^{(1)}}~~~~~\\
 + \nu
     \left\{c_1^4\left[\frac{(\nu+1)(\nu+2)(\nu+3)}{4!}\,{\cal D}^4
                     + \frac{3\,\nu^2+12\,\nu+11}{6}\,{\cal D}^3
                     - \frac{\nu^2+2\,\nu}{2}\,{\cal D}^2
                     - \frac{3\,\nu+5}{2}\,{\cal D}\right.\right.
 ~~~~~~\ \nonumber\\ 
                 \left.\left.
                     + \frac{3\,\nu+4}{6}
                 \right]
         + \,c_1^2c_2\,(\nu+2)
                 \left[\frac{\nu+3}{2}\,{\cal D}^2
                    + {\cal D}
                    - 1 
                 \right]
         + c_2^2\,\frac{(\nu+4)(3\nu+7)}{6}
      \right\}           
         {{\cal A }_{\nu+4}^{(1)}\choose{\mathfrak A}_{\nu+4}^{(1)}}
          \label{eq:image.u.4.nu}~~~~~\\
 +\,{\cal O}\left({\cal D}^{5}\,{{\cal A }_{\nu+4}^{(1)}
  \choose{\mathfrak A}_{\nu+5}^{(1)}}
            \right)\,.
~~~~~~~~~~~~~~~~~~~~~~~~~~~~~~~~~~~~~~~~~~~~~~~~~~~~~~~~~~~~~~~~~~~~~~~~~~~~\nonumber
\end{eqnarray}
\end{subequations}
%Eqs (C2a) (C2b) (C2c)

\textbf{2.} The images of the coupling accompanied by logarithms of the
momentum, i.e., $\left(a_{(2)}\right)^\nu L$, in accordance with Eq.\ 
(\ref{eq:a2Ltoa1}), are
\begin{eqnarray}
 {{\cal L}_\nu \choose {\mathfrak L}_\nu}
\equiv
 {\textbf{A}_\text{E}\choose \textbf{A}_\text{M}}
 \left[\left(a_{(2)}\right)^\nu L\right]
 &=&
 { {\cal A }_{\nu-1}^{(2)} \choose {\mathfrak A}_{\nu-1}^{(2)}}
  + c_1\,{\cal D}\,{ {\cal A }_{\nu}^{(2)}
  \choose {\mathfrak A}_{\nu}^{(2)}}
  - c_1^2\,{ {\cal A }_{\nu+1}^{(2)}
  \choose {\mathfrak A}_{\nu+1}^{(2)}}+\frac{c_1^3}{2}\,
  { {\cal A }_{\nu+2}^{(2)}
  \choose {\mathfrak A}_{\nu+2}^{(2)}}
  \nonumber \\
 &+& {\cal O}\left({\cal A}_{\nu+3}^{(2)}\right)\,.
\label{eq:image-a2-rho}
\end{eqnarray}
%Eq (C3)
Following Eq.\ (\ref{eq:A8.3L}), this leads to
\begin{eqnarray}
  {{\cal L}_\nu^{(3)} \choose {\mathfrak L}_\nu^{(3)}}
  = {{\cal A }_{\nu-1}^{(1)} \choose {\mathfrak A}_{\nu-1}^{(1)}}
  + c_1\,\nu~{\cal D}\,{{\cal A}_{\nu}^{(1)}
     \choose {\mathfrak A}_{\nu}^{(1)}}
  + \nu\left[c_1^2\,
             \left(\frac{\nu+1}{2!}\,{\cal D}^2
                + {\cal D} - 1
             \right)
           + c_2  
        \right]
        {{\cal A }_{\nu+1}^{(1)} \choose {\mathfrak A}_{\nu+1}^{(1)}}
 ~~~~~~~~~\nonumber \\
  +\,c_1\,\nu
     \left\{c_1^2\left[\frac{(\nu+1)(\nu+2)}{3!}\,{\cal D}^3
                        + \frac{2\,\nu+3}{2}\,{\cal D}^2
                        + (\nu+1)\,{\cal D}
                        - \frac12
                  \right]
           + c_2(\nu+2)\,{\cal D}
     \right\}
         {{\cal A }_{\nu+2}^{(1)}\choose{\mathfrak A}_{\nu+2}^{(1)}}
 \nonumber \\
  +\,{\cal O}
      \left({\cal D}^{3}{{\cal A}_{\nu+3}^{(1)}
      \choose {\mathfrak A}_{\nu+3}^{(1)}}
      \right)\,.
 ~~~~~~~~~~~~~~~~~~~~~~~~~~~~~~~~~~~~~~~~~~~~~~~~~~~~~~~~~~~~~~~~~~~~~~~~~~~~~~~~~~~      
 \label{eq:B8}
\end{eqnarray}
%Eq (C4)

%%%%%%%%%%%%%%%%%%%%%%%%%%%%%%%%%%%%%%%%%%%%%%%%%%%%%%%%%%%%%%%%%%%%%%%
%%%%%%%%%%%%%%%%%%%%%%%%%%%%%%%%%%%%%%%%%%%%%%%%%%%%%%%%%%%%%%%%%%%%%%%
\section{Representation of the D-function}
%%%%%%%%%%%%%%%%%%%%%%%%%%%%%%%%%%%%%%%%%%%%%%%%%%%%%%%%%%%%%%%%%%%%%%%

\textbf{1.}
The first three coefficients $d_1, d_2,d_3$ for the $D_\text{S}$
function of two scalar quark currents,
\begin{eqnarray}
 D_\text{S}(Q^2)
  &=& 3\,m_b^2(Q^2)
       \left[1+\sum_{n>0} d_n
        \left({\alpha_s(Q^2)\over\pi}\right)^n
        \right]\, ,
  \label{dn}
\end{eqnarray}
%Eq (D1)
have been calculated in \cite{Che96} and read
\begin{subequations}
\begin{eqnarray}
 d_1 &=& C_\text{F}
           \left[\frac{17}{4}\right]\,,\\
 d_2 &=& C_\text{F}^2
          \left[\frac{691}{64} 
                  - \frac{9}{4}\,\zeta(3)
          \right]
      +\,C_\text{F}\,C_\text{A}
          \left[\frac{893}{64} 
                  - \frac{31}{8}\,\zeta(3)
          \right]
      +\,T_\text{R}\,N_f\,C_\text{F}
          \left[-\frac{65}{16} 
                  + \zeta(3)
          \right]\,,~~
\end{eqnarray}
%Eqs (D2a) (D2b)
\begin{eqnarray}
 d_3 &=& C_\text{F}^3
          \left[\frac{23443}{768} 
                 - \frac{239}{16}\,\zeta(3)
                 + \frac{45}{8}\,\zeta(5)
          \right]
      +\,C_\text{F}^2\,C_\text{A}
          \left[\frac{13153}{192} 
                 - \frac{1089}{32}\,\zeta(3)
                 + \frac{145}{16}\,\zeta(5)
          \right]
\nonumber\\
     &+& C_\text{F}\,C_\text{A}^2
          \left[\frac{3894493}{62208} 
                 - \frac{2329}{96}\,\zeta(3)
                 + \frac{25}{48}\,\zeta(5)
          \right]
\nonumber\\
     &+& T_\text{R}\,N_f\,C_\text{F}^2
          \left[-\frac{88}{3}  
                 + \frac{65}{4}\,\zeta(3)
                 + \frac{3}{4}\,\zeta(4)
                 - 5\,\zeta(5)
          \right]
\nonumber\\
     &+& T_\text{R}\,N_f\,C_\text{F}\,C_\text{A}
          \left[-\frac{33475}{972} 
                 + \frac{22}{3}\,\zeta(3)
                 - \frac{3}{4}\,\zeta(4)
                 + \frac{5}{6}\,\zeta(5)
          \right]
\nonumber\\
     &+& T_\text{R}^2\,N_f^2\,C_\text{F}
          \left[\frac{15511}{3888} 
                 - \zeta(3)
          \right]\,.
 \label{RS}
\end{eqnarray}
%Eq (D2c)
The result for $d_4$ was obtained quite recently in \cite{BCK05},
\begin{eqnarray}
 d_4 &=& N_f^3
            \left[-\frac{520771}{559872}
                  + \frac{65}{432}\,\zeta(3)
                  + \frac{1}{144}\,\zeta(4)
                  + \frac{5}{18}\,\zeta(5)
            \right]
\nonumber\\
     &+& N_f^2
            \left[\frac{220313525}{2239488}
                  - \frac{11875}{432}\,\zeta(3)
                  + \frac{5}{6}\,\zeta^2(3)
                  + \frac{25}{96}\,\zeta(4)
                  - \frac{5015}{432}\,\zeta(5)
            \right]
\nonumber\\
     &+& N_f\left[-\frac{1045811915}{373248}
                  + \frac{5747185}{5184}\,\zeta(3)
                  - \frac{955}{16}\,\zeta^2(3)
                  - \frac{9131}{576}\,\zeta(4)
                  + \frac{41215}{432}\,\zeta(5)
                   \right. \nonumber \\
& & \phantom{N_f^0}
               \left.~
                  + \frac{2875}{288}\,\zeta(6)
                  + \frac{665}{72}\,\zeta(7)
            \right] \nonumber\\
     &+& N_f^0 \left[\frac{10811054729}{497664}
                  - \frac{3887351}{324}\,\zeta(3)
                  + \frac{458425}{432}\,\zeta^2(3)
                  + \frac{265}{18}\,\zeta(4)
                  + \frac{373975}{432}\,\zeta(5)
               \right. \nonumber \\
     & & \phantom{N_f^0}
               \left.~
                  - \frac{1375}{32}\,\zeta(6)
                  - \frac{178045}{768}\,\zeta(7)
               \right]\,.
\label{NPsip*QQ}
\end{eqnarray}
\end{subequations}
%Eq (D2d)

\textbf{2.} The coefficients $\gamma_i$ determine the expansion of the
quark-mass  anomalous dimension in analogous manner as Eq.\ 
(\ref{eq:betaf}) does with respect to the expansion of the
$\beta$-function; viz.,
\begin{eqnarray}
 \frac{d}{dL}\ln\left(m(L)\right)
  = - \sum_{i \geq 0} 
  \gamma_i\left(\frac{\alpha_{s}(L)}{4 \pi}\right)^{i+1},
\label{eq:gammaf}
\end{eqnarray}
%Eq (D3)
with the following expressions, see \cite{Che97},
\begin{eqnarray}
 \gamma_0
  &=& 3 C_\text{F}\,;
   \label{g0}\\
 \gamma_1
  &=& \bigg[\frac{202}{3}-\frac{20}{9}\,N_f
       \bigg]\,;
   \label{g1}\\
 \gamma_2
  &=& \bigg[1249
            -\bigg(\frac{2216}{27}+\frac{160}{3}\zeta(3)\bigg)\,N_f
            -\frac{140}{81}\,N_f^2
       \bigg]\,;
   \label{g2}\\
 \gamma_3
  &=& \bigg[\frac{4603055}{162}
            +\frac{135680}{27}\zeta(3)
            -8800\zeta(5)
       \nonumber\\
  & &{}     -\bigg(\frac{91723}{27}
                  +\frac{34192}{9}\zeta(3)
                  -880\zeta(4)
                  -\frac{18400}{9}\zeta(5)
             \bigg)\,N_f
        \nonumber\\
  & &{}     +\bigg(\frac{5242}{243}
                  +\frac{800}{9}\zeta(5)
                  -\frac{160}{3}\zeta(4)
             \bigg)\,N_f^2
            -\bigg(\frac{332}{243}
                  -\frac{64}{27}\zeta(3)
             \bigg)\,N_f^3
       \bigg]\,,
  \label{g3}
\end{eqnarray}
%Eqs (D4) (D5) (D6) (D7)
where $\zeta(\nu)$ denotes the Riemann $\zeta$ function.
\end{appendix}

\end{document}